\documentclass[aps,pra,10pt,twocolumn,nofootinbib,floatfix,superscriptaddress]{revtex4-2}
\usepackage[a4paper,left=1.8cm,right=1.8cm,top=2.5cm,bottom=3cm]{geometry}

\usepackage[utf8]{inputenc}
\usepackage[T1]{fontenc}
\usepackage{amssymb}
\usepackage{graphicx}
\usepackage[dvipsnames]{xcolor} 
\usepackage{amsmath,bm}
\usepackage{upgreek}
\usepackage{appendix}

\usepackage{pgfplots}
\DeclareUnicodeCharacter{2212}{−}
\usepgfplotslibrary{groupplots,dateplot}
\usetikzlibrary{patterns,shapes.arrows}
\pgfplotsset{compat=newest}

\begin{document}


\title{Enhanced Atom Capture via Multi-Frequency Magneto-Optical Trapping}

\author{Benjamin Hopton}
\address{School of Physics and Astronomy, University of Nottingham, University Park, Nottingham, NG7 2RD, UK} 
\author{Alexander Abbey}
\address{School of Physics and Astronomy, University of Nottingham, University Park, Nottingham, NG7 2RD, UK} 
\author{David Johnson}
\address{School of Physics and Astronomy, University of Nottingham, University Park, Nottingham, NG7 2RD, UK} 
\author{Daniele Baldolini}
\address{School of Physics and Astronomy, University of Nottingham, University Park, Nottingham, NG7 2RD, UK} 
\author{Matt Overton}
\address{School of Physics and Astronomy, University of Nottingham, University Park, Nottingham, NG7 2RD, UK} 
\author{Nathan Cooper}
\address{School of Physics and Astronomy, University of Nottingham, University Park, Nottingham, NG7 2RD, UK} 
\author{Joseph Aziz}
\address{Department of Physics, Royal Holloway, University of London, Egham Hill, Egham TW20 0EX, UK}
\author{Richard Howl}
\address{Department of Physics, Royal Holloway, University of London, Egham Hill, Egham TW20 0EX, UK}
\author{Lucia Hackerm\"{u}ller}
\address{School of Physics and Astronomy, University of Nottingham, University Park, Nottingham, NG7 2RD, UK}


\vspace{10pt}

\begin{abstract}
Magneto-optical traps are central to atomic and molecular quantum technologies and precision tests of fundamental physics, where both sensitivity and bandwidth scale strongly with atom number and loading rate. We demonstrate that employing multiple, closely spaced optical frequency components in the cooling light of a $^{87}$Rb magneto-optical trap---without utilizing any additional slowing techniques---can double the steady state atom number and increase the loading rate by up to a factor of 4, compared to a conventional single-frequency implementation. Subsequently, we capture up to $1.0(1)\times10^{10}$ atoms with a loading rate of up to $1.3(2)\times 10^{11}\,\mathrm{atoms\,s^{-1}}$ from a thermal background.
Numerical simulations reproduce the observed trends and predict substantially larger gains for increased trap sizes beyond our experimental bounds. By re-examining earlier studies of multi-frequency atom capture in the context of modern experimental hardware and emerging applications, we show that previously identified limitations can be avoided and establish multi-frequency cooling as a practical and scalable route to high-flux cold-atom sources. These results have immediate applications in portable atom-based quantum sensing, where higher bandwidth and precision can be achieved without forgoing compactness, and in atom-interferometric tests of fundamental physics, which benefit from access to larger-mass quantum systems. 
\end{abstract}

\maketitle

\section{Introduction}
%
The magneto-optical trap (MOT) is an essential underpinning technology for the creation of cold atom clouds \cite{Raab1987} and the generation of Bose--Einstein condensates (BECs) or other ultracold gases \cite{Georgescu2020,UFG}. Already a vital research tool, the MOT is now also a critical component of many emerging quantum technologies (QTs), including gravimeters \cite{gravimeter2,Cassens2025}, accelerometers \cite{portaccelerometer,Meng2024}, magnetometers \cite{MIT,coldmag}, atomic clocks \cite{Kobayashi2024,Chu2025} and neutral-atom quantum simulators \cite{solvent_qsim,qsim_spin_interactions}. All of these technologies stand to benefit by higher sensitivity and improved bandwidth from the ability to capture more atoms at a higher rate. Additionally, a raft of recent theoretical proposals employ cold atoms---usually via atom interferometry \cite{Bongs2019}---to address key unsolved questions in fundamental physics. Examples include searches for light dark matter and gravitational waves \cite{AION,gravwav,gravwav2,gravdark}, precision tests of the equivalence principle \cite{equivtest0,equivtest} and searches for variation in fundamental constants \cite{alphadrift}, as well as experimental tests of wave function collapse models \cite{borntest,Schrinski2023} and quantum gravity \cite{qgrav1,Howl2023}. 

A key challenge for all such endeavors is to maximize experimental sensitivity while minimizing the influence of systematic biases and long-term drifts. Increasing the total number of captured atoms, and the repetition rate of the experiment, is thus of central importance; measurement sensitivity per unit operating time scales as the square root of the repetition rate---vital for QT sensor limitations---and absolute sensitivity often scales more favorably than this with the number of atoms employed per cycle \cite{Schrinski2023,Howl2023,Buchmueller2023}. Fundamental research into the properties of antimatter \cite{antihydrogen,positronium1,positronium2} and ultracold molecules \cite{molcool1,molcool2} also brings new incentives for efficient particle capture in challenging circumstances.

The above considerations provide a strong motivation for enhanced atom capture, while advances in experimental hardware since the early days of the MOT have expanded the range of tools available to experimentalists. As a result, there is resurgent interest in optimizing magneto-optical atom capture. Previous successful demonstrations of enhanced atom capture in a MOT were limited to two optical frequency components and obtained no more than a doubling of the loading rate \cite{Sinclair1994,twocolour_2012}. Although a significant improvement in the capture of Li atoms from a beam source was demonstrated using multiple optical frequencies \cite{Li_beam_MOT}, and more modest improvements were obtained with 2D MOTs \cite{LiMF2DMOT,KMF2DMOT}, attempts to transfer the technique to other atomic species and more standard MOT configurations were largely unsuccessful \cite{Gibble1992,MFMOT_fail}, and a number of pitfalls were identified \cite{f_v_reversal,collision_loss}. 

We critically examine these pitfalls and devise techniques for circumventing them. Implementing several such techniques, we more than tripled the loading rate of a $^{87}$Rb MOT to $1.3(2)\times 10^{11}$\,s$^{-1}$. With a $1/e^2$ beam waist radius of 15\,mm (truncated at 19\,mm) the maximum steady-state atom number is approximately doubled, reaching $1.0(1)\times 10^{10}$ atoms, with characteristic MOT loading time constants as short as $\tau=0.1\,\mathrm{s}$. 
These results clearly demonstrate the potential for known drawbacks of multi-frequency atom capture to be circumvented through the techniques described herein. As such, they pave the way for much larger improvements in atom capture in 
very large MOTs \cite{Camara2014VLMOT}, 2D MOTs and linear beam slowers \cite{LiMF2DMOT,KMF2DMOT,Li_beam_MOT,hoffnagle_clock}, or with very light particle species \cite{antihydrogen,positronium1,positronium2,hydrogen}. 
Application of multi-frequency atom capture within these contexts holds great promise for future tests of fundamental physics, and will also benefit a wide range of emerging QTs.

\section{Enhanced atom capture}

\begin{figure}
    \centering
    \includegraphics[width=\linewidth]{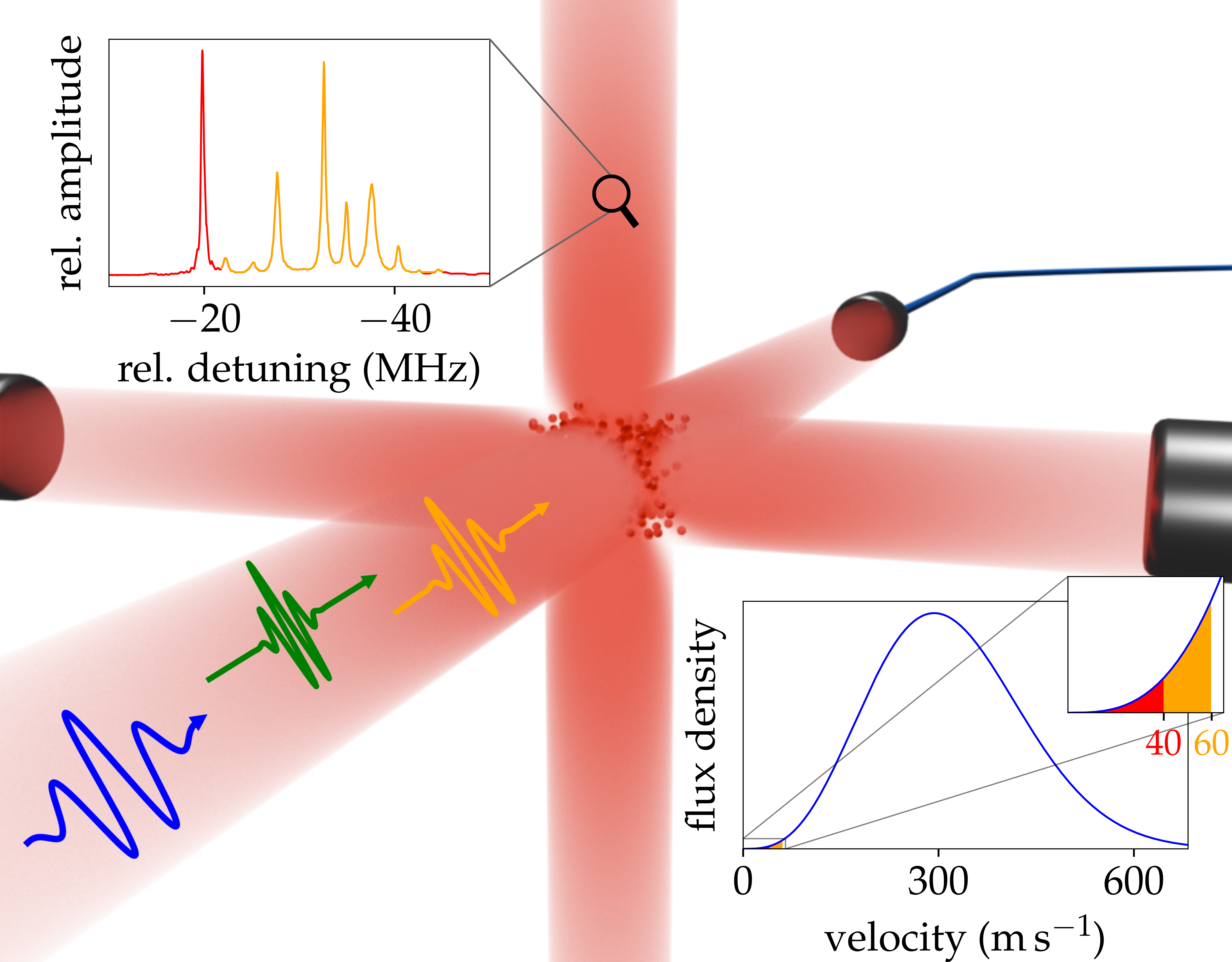}
    \caption{Schematic illustrating the implementation and principles behind the six-beam multi-frequency MOT used for enhanced atom capture. \emph{Top left:} Example optical frequency spectrum, measured with a Fabry--Perot interferometric analyzer, used for enhanced capture. The conventional trapping frequency is shown in red, while the additional multi-frequency cooling components are shown in amber. \emph{Bottom right:} Velocity distribution of $^{87}$Rb atomic flux density at room temperature. The red region shows the proportion of atomic flux with a velocity $<40\,\mathrm{m\,s^{-1}}$ (that of a typical MOT), while the amber region illustrates the additional trappable flux with a capture velocity increase to $60\,\mathrm{m\,s^{-1}}$.
    This illustrates the scale of the benefits available from an increased capture velocity through multi-frequency cooling.}
    \label{fig:fig1}
\end{figure}

Characteristic thermal speeds of trappable atomic species at room temperature are typically hundreds of metres per second (or higher), leading to Doppler shifts of hundreds of megahertz. The natural linewidths of atomic transitions used for Doppler cooling and magneto-optical trapping are typically on the order of 5\,MHz. A single, narrow-linewidth laser therefore addresses only a tiny fraction of the available atomic velocity classes (see fig. \ref{fig:fig1}). 
Power broadening and Zeeman-assisted slowing in the MOT's magnetic field gradient partially mitigate this problem, but still typical capture velocities remain at only 10--40\,m\,s$^{-1}$ for most MOTs. For a room-temperature gas of rubidium-87 atoms and a capture velocity of 30\,m\,s$^{-1}$, this leads to approximately 1 in $10^5$ of the incident atoms being captured. 

Multi-frequency magneto-optical trapping employs laser beams that contain multiple frequency components, spaced on the order of the natural linewidth apart, allowing a wide range of atomic velocities to be addressed simultaneously. This allows for more continuous cooling: as the atom decelerates and its Doppler shift changes, it becomes resonant with a neighbouring laser frequency after shifting off-resonance with the previous one. A simple upper limit on the performance of multi-frequency atom capture can be found by assuming that atoms are decelerated at the maximum possible rate, $a_\mathrm{max}=\hbar k\Gamma/2m$, across the entire trapping region. This suggests that very large increases in loading rate, by 2 or even 3 orders of magnitude, might be possible using multi-frequency light under ideal circumstances. More than this cannot be achieved unless exploiting coherent effects that require more complex hardware, such as amplified Doppler cooling \cite{amplifiedDopplercooling} or the bichromatic force \cite{bichromatic_sims,bichromatic_li}. 

Real systems lie between this bound and the performance of a single-frequency MOT. To assess likely real performance, we developed a theoretical model of atomic deceleration and capture and simulated the process using the PyLCP software suite \cite{PyLCP}. Full details of the model are given in Methods, and selected results are shown in figs. \ref{fig:vs B-field grad} and \ref{fig:vs beam diam}.  For conditions matching those of our experiment, these simulations predict a three-fold increase in loading rate and a doubling of the steady state atom number. Extending the model to beam diameters beyond our experimental limitations, we find that for conditions such as those used in \cite{Camara2014VLMOT} and appropriate multi-frequency spectra, the model predicts more than an order of magnitude enhancement of the loading rate through the use of multi-frequency cooling light. The primary factor driving this improvement for larger MOTs is the available deceleration distance; as such, even greater performance improvements are likely possible in systems loaded from 2D MOTs, Zeeman slowers and similar atom capture apparatus \cite{zeeman_asaf,zeeman_2025,2dmot_2006,2dmot_2007,2dmot_2009}.

Multi-frequency magneto-optical trapping therefore has significant potential to enhance experiments based on MOTs by increasing the rate at which atoms can be captured and, correspondingly, the maximum number of atoms that can be trapped. However, two major obstacles were identified to its use \cite{MFMOT_fail,f_v_reversal,collision_loss}. A key component of this work is the development of techniques to circumvent these obstacles and enable successful deployment. Below, we briefly discuss each obstacle and how it can be overcome.

Firstly, multi-frequency cooling light must contain components detuned tens to hundreds of megahertz below relevant atomic transitions. This has been predicted \cite{GallagherPritchard1989} and shown \cite{collision_loss,Hoffman1994collisional_losses} to induce a higher rate of fine-structure changing collisions between trapped atoms, which are a loss channel for the MOT. We circumvent this obstacle by applying multi-frequency cooling light in a ring-shaped beam generated using a pair of axicons (see fig \ref{fig:axicon beam real and sim}); the multi-frequency light thus slows atoms incident upon the exterior of the trapping region but has minimal intensity overlap with the trapped atoms. A superior alternative, avoided herein solely for reasons of economy, would be to use multi-frequency cooling light tuned to the D\textsubscript{1} line of the relevant atomic species, in combination with single-frequency trapping light tuned to the D\textsubscript{2} line; the D\textsubscript{2} trapping light would also provide repumping to avoid dark-state pumping from the D\textsubscript{1} light. 

Secondly, the multi-frequency light employed has traditionally been generated by phase modulation. Phase modulation yields an additional time-varying phase term, and the Jacobi-Anger identity can be used to show that such light can be decomposed into a number of separate frequency components or ``sidebands'', spaced by the modulation frequency. However, this remains by definition equivalent to a time-varying phase/frequency, and the Jacobi-Anger identity defines fixed amplitude and phase relationships between the components. The resulting frequency spectra led to a reversal of the velocity dependence of the optical force on the atoms at low atomic velocities, resulting in heating instead of cooling \cite{f_v_reversal}. This reduced the resulting capture efficiency. Theoretical studies showed that such heating effects occurred only when light was present that was blue-detuned with respect to the atomic transition \cite{f_v_reversal}. We therefore employ a single-sided multi-frequency spectrum with a sharp cutoff, such that negligible optical power is blue-detuned with respect to the atomic transition. We generate this using one of two approaches explained in detail in Methods. Both depend on acousto-optic, rather than electro-optic, modulation; this avoids the symmetric spectral form yielded by standard phase modulation techniques. 
The top left inset of fig. \ref{fig:fig1} shows an example of the kind of frequency spectrum that is thus generated. 

These two experimental alterations sufficiently mitigate the deleterious effects previously observed as to make multi-frequency magneto-optical trapping significantly advantageous over a single-frequency approach. This is demonstrated by the results below.

\section{Experimental setup}
The multi-frequency MOT is built upon the traditional $^{87}$Rb MOT setup: we use three pairs of counter-propagating---not retro-reflected---Gaussian beams (38\,mm diameter with a $1/e^2$ radius $w\sim 15$\,mm), consisting of cooling light tuned to $\delta\sim-20\,\mathrm{MHz}$ from the $F=2\rightarrow F'=3$ transition, and repumper light tuned on-resonance with the $F=1\rightarrow F'=2$ transition. For purposes of naming conventions and clarity, the Gaussian beam cooling light in the typical MOT set up will be referred to as the `trapping' light. The powers in each beam were $P_{\mathrm{trap.}}\approx 20\,\mathrm{mW}$ and $P_\mathrm{repump.}\approx18\,\mathrm{mW}$. Then we overlapped a ring-shaped beam---created via a pair of axicon lenses---with the traditional MOT beams, with $\sim 5$ additional further red-detuned frequencies present within (see fig. \ref{fig:MF FP peaks}), tuned to $\delta_i\sim -25\,\mathrm{MHz}\text{ to }-45\,\mathrm{MHz}$ from the $F=2\rightarrow F'=3$ transition. For convention this light is referred to as the `cooling' light. Typically $P_\mathrm{cool.}\approx45\,\mathrm{mW}$ per beam. Anti-Helmholtz coils are used to produce the magnetic field, achieving field gradients up to $\sim22\,\mathrm{G\,cm^{-1}}$.

We use absorption imaging to record values of atom number in our experiment. A laser pulse of duration $\sim35\,\upmu\mathrm{s}$, tuned on-resonance to the $F=2\rightarrow F'=3$ transition, is fired to the atom cloud a short fall time of $\sim 15\,\mathrm{ms}$ after the cooling and trapping beams and the magnetic field were turned off. Experimentally it was found that a short period of $\sim20\,\mathrm{ms}$ at the end of loading where only single frequency trapping light was present yielded more optimal results. The standard absorption imaging technique is used to construct a two-dimensional projection of the MOT's density distribution from its optical depth. This projection is then fit to a two-dimensional Gaussian and rescaled to represent its true physical extent, since the extent of the atom cloud expanded beyond the field of view of the imaging beam. Integration of this rescaled density distribution then gives the MOT's atom number.

We obtained values of loading rate and steady state atom number by conducting scans where the MOT load time was varied at a certain point in parameter space and fitting the resulting averaged recorded atom number to the equation 
\begin{equation}
    N = R\tau\left( 1-e^{-t/\tau}  \right),
\label{loadrate}
\end{equation}
where $R$ is the load rate and $\tau$ is the characteristic MOT load time, from which the load rate and steady state atom number $N_\mathrm{ss}=R\tau$ is calculated and extracted. One value of loading rate and steady state atom number is taken per load time scan (see insets in fig \ref{fig:vs B-field grad}; each plot produces a single data point). To mitigate inaccuracies in shutter timings coinciding perfectly with $t=0\,\mathrm{s}$, an offset on the time axis was allowed to be added to the fitting, rather than forcing the fit through the origin, to better represent the rate of loading.

\section{Results}
Experimental results are shown in figs. \ref{fig:vs B-field grad} and \ref{fig:vs beam diam}, together with PyLCP simulations. The key aspect in which it differs from a conventional MOT system is in the inclusion of a ring-shaped set of `cooling' beams containing multiple optical frequency components, in addition to the traditional Gaussian `trapping' beam of a single frequency MOT; the intensity profile and frequency spectrum of these additional beams are given in fig. \ref{fig:MF FP peaks}.

Figs. \ref{fig:vs B-field grad} and \ref{fig:vs beam diam} show the resulting improvements in loading rate and steady state atom number under different experimental conditions. These improvements exceed those previously reported for dual-frequency MOTs \cite{twocolour_1994,twocolour_2012}, with the absolute recorded load rates being notably high, particularly for the rubidium-87. This is important for high-bandwidth sensing and high-precision fundamental physics experiments. The discrepancy between simulation and experiment at low magnetic field gradient owes to the lack of adequate trapping forces for small field gradient; with only very small trapping forces minor experimental imperfections such as beam power imbalance etc., together with small effects not included in the simulation model (such as the radiation pressure force resulting from light spontaneously emitted by other atoms in the cloud), can prevent effective trapping. 

\begin{figure}  
\includegraphics[width=\linewidth]{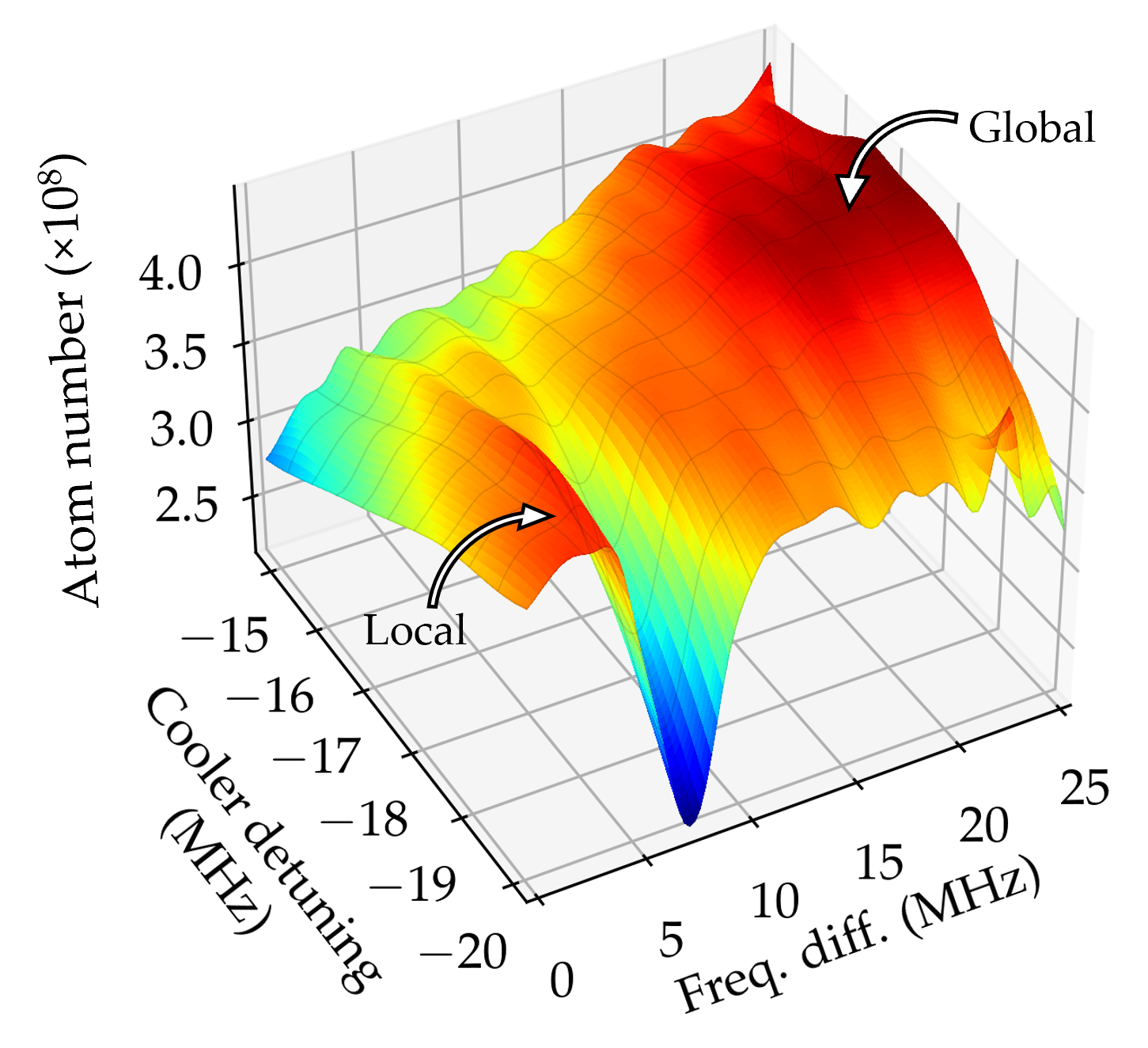}
\caption{Surface plot showing number of atoms loaded in a dual-frequency MOT after 5\,s load time as a function of both fundamental frequency detuning and dual-frequency spacing, where a surface interpolation has been used for better visual representation. A distinct trough can be seen across a range of fundamental frequency detunings for a dual-frequency spacing $\sim$7\,MHz. A local maximum in atom number, uncovered in previous work \cite{twocolour_2012,timthesis}, exists at a smaller dual-frequency spacing than this, while the global maximum is at significantly higher dual-frequency spacing $\sim$20\,MHz.}
\label{fig:trough}
\end{figure}

\begin{figure}  
    \includegraphics[width=\linewidth]{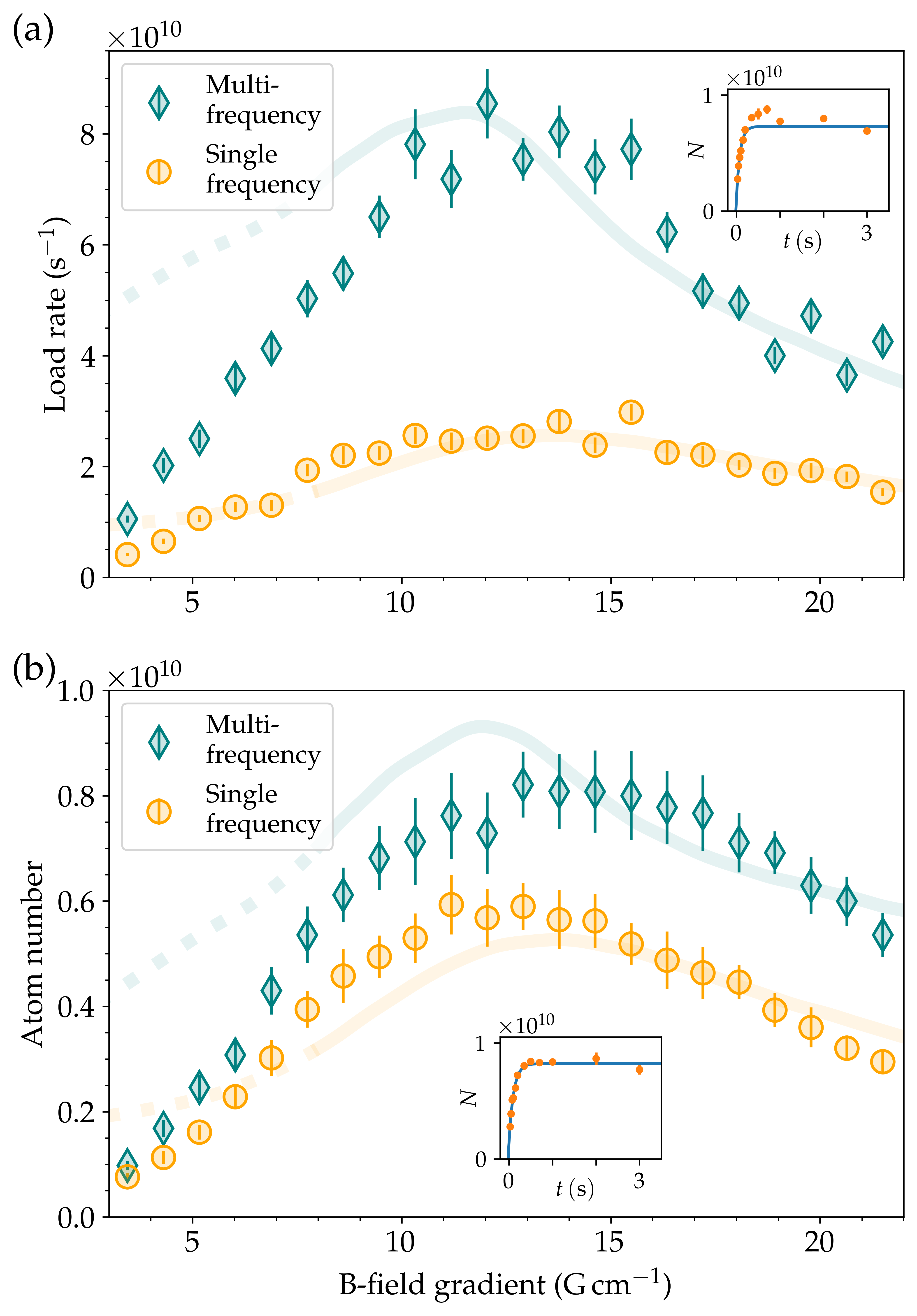}
    \caption{Loading rate (a) and steady-state atom number (b) for single frequency and multi-frequency magneto-optical traps as a function of axial magnetic field gradient, under otherwise identical conditions. Each data point represents the value obtained from a loading curve fit (see insets), with error bars calculated from the square root of the covariance matrix produced. The inset in (a) shows the multi-frequency loading curve for the data point at 12.0\,$\mathrm{G\,cm^{-1}}$, producing a load rate of $8.5(6)\times 10^{10}\,\mathrm{s^{-1}}$. The inset in (b) shows the multi-frequency loading curve for the data point at 12.9$\,\mathrm{G\,cm^{-1}}$, producing a steady state atom number of $8.2(6)\times 10^{9}$. Solid lines show corresponding numerical simulations in line with experimental conditions. Dotted lines indicate regions where the model is not applicable (see text). Further information is found in the appendix, sections \ref{additional experimental info} and \ref{simulation info}.}
    \label{fig:vs B-field grad}
\end{figure}

\begin{figure}
    \centering
    \includegraphics[width=\linewidth]{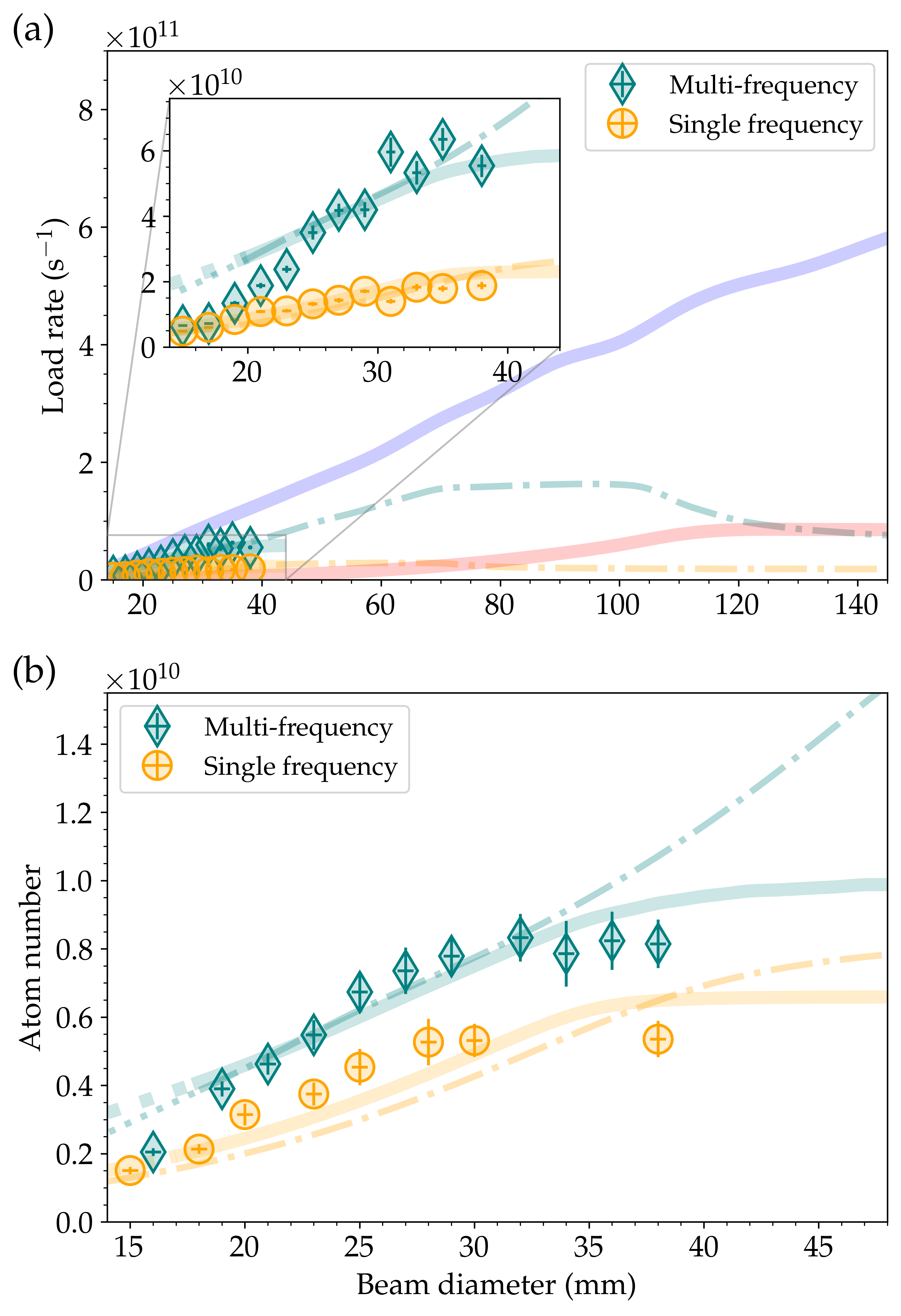}
    \caption[Caption for LOF]{(a) Loading rate and (b) steady state atom number for single frequency and multi-frequency magneto-optical traps as a function of MOT beam diameter. Data points are extracted from loading curve fits for given experimental conditions at 12.9 $\mathrm{G\,cm^{-1}}$. Vertical error bars are obtained from these fits, with horizontal error bars indicating the estimated $1\sigma $ uncertainty in the beam truncation measurement. Solid green (multi-frequency) and orange (single frequency) lines show numerical simulations modelling beam truncation consistent with our experiment. Dash-dotted lines instead vary the beam waist and include additional frequency components, providing a continuation to our experimental conditions. Solid blue (multi-frequency) and red (single frequency) lines show simulations that extend further beyond our experimental regime, illustrating the performance gains to be had at larger beam diameters. Dotted segments indicate where the model deviates from experiment. Further information is found in the appendix, sections \ref{additional experimental info} and \ref{simulation info}.}
    \label{fig:vs beam diam}
\end{figure}

Fig. \ref{fig:trough} shows how the loading rate of a MOT generated with two cooling frequencies varies as a function of those two frequencies. It reveals two distinct regimes in which loading is substantially enhanced: the global maximum that we target in this work, based on Doppler cooling of an increased spread of velocities, and a local maximum with much smaller frequency difference between the two optical frequencies. This local maximum was the condition used in some previous studies such as \cite{twocolour_2012}, and is likely attributable to coherent bichromatic forces \cite{twocolour_2012,bichromatic_sims}. 

Fig. \ref{fig:vs beam diam} shows how simulations indicate that multi-frequency MOTs become more advantageous as the physical size of the trap is increased. For example, with 100\,mm trapping beams as employed in \cite{Camara2014VLMOT}, our simulations predict up to 7-fold increase in loading rate, with up to an order-of-magnitude improvement for even larger diameters. The agreement between these simulations and our measurements within the experimentally tested region adds credibility to the simulations' predictions elsewhere. The trend towards increased benefit at larger MOT size is easily understandable as the result of the increased deceleration distance available---when the limit imposed on capture velocity by the stopping distance is increased, the ratio between this and the limit imposed in single-frequency MOTs by the frequency width of the atomic response becomes greater. This implies that devices such as linear slowers or 2D MOTs, which can offer much larger deceleration distances \cite{hofnagle_slower,zeeman_asaf,zeeman_2025,2dmot_2006,2dmot_2007,2dmot_2009}, stand to benefit particularly strongly from the use of multi-frequency cooling light. Appropriate use of the technique in combination with such devices may allow a range of fundamental physics tests with comparatively accessible cold atom experiments \cite{fundamentals,space_fundamentals}. 

\section{Applications}

An immediate application of high loading rate MOTs will be in high bandwidth quantum sensing. This will be crucial to the next generation of atomic gravimeters, cold-atom magnetometers and accelerometers. Improving the load rate can directly reduce a sensor's ``dead time'', increasing bandwidth while reducing aliased phase noise \cite{Cheinetinfsens}, making low load times desirable for portable interferometer systems. Alternatively, zero dead time sensors would generally benefit from an increased sensitivity due to higher atomic flux \cite{Xuecontiousint} or more atoms loaded within time restrictions \cite{Duttacontinuous,Hansel2001}. Recapture schemes also benefit from either increased sensitivity or increased operating rate \cite{McGuinness2012}, as will many cold-atom quantum computing schemes \cite{Chew2022}. Combination of multi-frequency trapping with non-magnetic MOT equivalents, such as those described \cite{troop,moptrap}, may prove particularly effective for high-bandwidth quantum sensors by eliminating the need for magnetic coils, together with their associated challenges around SWAP, heat dissipation and switching times.

Standard signal-to-noise considerations mean that, for given experimental resources, sensitivity scales as the square root of either the number of atoms loaded in a MOT or the repetition rate. 
As multi-frequency magneto-optical trapping also improves the steady state atom number, sensing methodologies that beat the standard quantum limit benefit more strongly. In particular, squeezed states can allow magnetometers \cite{Muessel2014}, clocks \cite{Robinson2024}  and interferometers \cite{Laudat2018} whose sensitivity scales linearly with atom number.

\begin{figure}
    \centering
    \includegraphics[width =\linewidth]{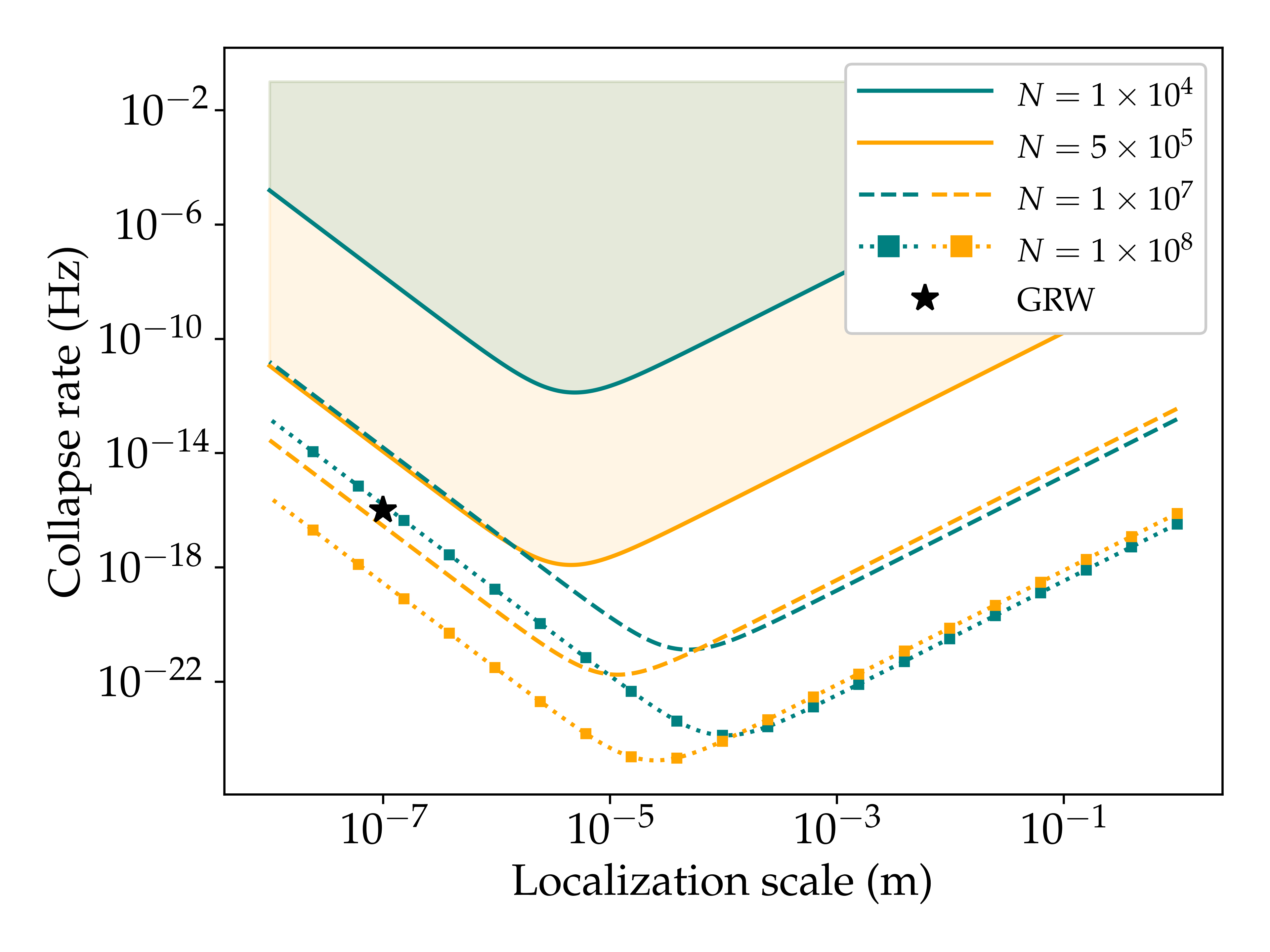}
    \caption{Projected bounds on the CSL parameter space from interferometry experiments with Rb BECs. These curves are calculated assuming the state of the art spin-echo interference protocols proposed in \cite{Schrinski2023}. Protocol $(I)$ assumes a single period of reversed dispersion (spin-echo) during the interference experiment. Protocol $(II)$ assumes an initial period of strong dispersion, followed by free evolution (weak dispersion) and then inverted strong dispersion. Each teal line assumes protocol $(I)$ with a density of $\rho = 1.65 \times 10^{-3}\, \text{kg}\,\text{m}^{-3}$, an interference time of $t = 100\,\text{ms}$, initial squeezing of $20 \,\text{dB}$ and a dispersion of $1.2\times 10^{-2}\,\text{Hz}$. Each yellow line assumes protocol $(II)$ with a density of $\rho = 2.25 \times 10^{-5} \,\text{kg}\,\text{m}^{-3}$, an interference time (free evolution time) of $t = 0.1\, \text{ms}$, initial squeezing of $30 \,\text{dB}$, a dispersion rate of $0.57 \,\text{Hz}$ and a strong dispersion rate of $56 \,\text{kHz}$ applied for $1 \,\upmu \text{s}$. The solid lines show the regions of the CSL parameter space that could be ruled out assuming the protocols and parameters values given in \cite{Schrinski2023}. In this work, the successful trapping of a cloud of $N \sim 10^{10}$ Rb atoms is demonstrated. From this we project a BEC atom number of $N \sim 10^7 \text{--} 10^8$. The dashed and dotted lines demonstrate the improvement on these bounds that would be gained with a BEC of $N \sim 10^7 \text{--} 10^8$, at the same densities respectively. The black star represents the original, theoretically motivated parameter values \cite{ghirardi1990markov}.} 
    \label{fig:CSL bounds}
\end{figure}

A more ambitious application is fundamental physics research. For certain experiments, much more favorable scaling is possible \cite{Schrinski2023,Howl2023,Buchmueller2023}; for example, \cite{Schrinski2023} proposes experimental methods yielding up to cubic dependence of experimental sensitivity on the number of atoms used. Given recent theoretical proposals that place cold atoms at or close to the forefront of currently available techniques for testing multiple areas of fundamental physics, any substantial increase in atom capture efficiency has the potential to make cold atom experiments very powerful tools in these areas. For example, it has been proposed to probe quantum gravity
with an atom interferometer, using either atoms directly from a MOT or a Bose--Einstein condensate \cite{Howl2023}. 
Furthermore, it has recently been demonstrated  that cold atom interferometry can be used to test theories with objective wave function collapse, such as continuous spontaneous localization (CSL) \cite{Schrinski2023} or fundamentally classical gravity \cite{oppenheim2023gravitationally}, potentially accessing large regions of parameter space. Assuming that the same enhancement in MOT atom numbers from multiple frequencies carries over to the BEC stage, we estimate that atom numbers on the order of $10^7$ $^{87}\mathrm{Rb}$ atoms could be achieved. If all of these atoms could used in the interferometer, this would significantly increase the region of CSL parameter space that could be experimentally ruled out---see Fig. \ref{fig:CSL bounds}. Alternatively, a MOT-based interferometric scheme could be employed, similar to \cite{Howl2023}, or a non-interferometric approach involving a single BEC or MOT,  similar to \cite{Howl2021}. 

Multi-frequency slowing and trapping is also particularly suited to trapping lighter particle species and is thus of interest for experiments with trapped antimatter, such as antihydrogen or positronium \cite{antihydrogen,positronium1,positronium2}. 

\section{Conclusions}

Multi-frequency trapping light has been used to increase the loading rate of a MOT by a factor $>3$, exceeding previous results for dual-frequency magneto-optical trapping. Simulations that correctly predict the trends in our experimental results also predict greater improvements for larger MOTs, reaching a factor of 12 in loading rate for a 75\,mm beam radius. The technique has applications in quantum sensing technologies, where it will offer significant improvements in bandwidth and sensitivity, as well as in a number of proposed tests of fundamental physics based on cold atoms. 





\section{Acknowledgements}
The experiment was supported by the grant 62420 from the John Templeton Foundation, the IUK project No.133086, EPSRC grants EP/T001046/1, EP/R024111/1, EP/M013294/1, EP/Y005139/1 and EP/Z533166/1, and by the European Commission grant ErBeStA (no.800942). We would also like to thank the developer team and contributors responsible for the production and maintenance of the \emph{labscript suite} \cite{STARKEY2019labscript,BILLINGTON2019labscript,10.1063/1.4817213labscript}, which was used to control the hardware of the experiment.


\appendix

\section{Additional experimental information}
\label{additional experimental info}
\subsection{Beam truncation}
Varying the beam size within a MOT modifies both the available capture volume and the intensity profile experienced by the atoms. However, changing the beam size by adjusting the beam waist is experimentally cumbersome, requiring significant and repeated modification of the optical setup. Such an approach is therefore impractical for systematic studies. Instead, to assess the performance of the multi-frequency MOT under realistic experimental constraints, we investigate its behaviour as a function of MOT beam \emph{truncation} diameter.

To systematically vary the effective beam diameter, adjustable irises are introduced into the beam path and used to impose a controlled truncation of the transverse beam profile. By progressively reducing the iris aperture, the outer regions of the beam are clipped, allowing a set of reproducible beam diameters to be realised under otherwise identical experimental conditions.

To ensure consistency, each iris is fixed in a region where the beam is well collimated and is carefully centered on the beam axis, with only the aperture diameter varied. The intensity profile of the beam is not changed to account for any power differences. In this way, the iris provides a simple and robust means of controlling the effective beam diameter, enabling a direct investigation of how MOT performance depends on the available beam size.

The quoted beam diameter then corresponds to the physically truncated beam diameter measured immediately prior to the chamber, rather than the intrinsic $1/e^2$ waist of a Gaussian beam. This provides a practical and experimentally accessible definition, as the transmitted beam is sharply bounded by the iris aperture. This method is particularly advantageous when working with non-Gaussian beam profiles, such as the ring-shaped beam produced by the axicon pair, for which no unique or natural definition of beam diameter exists.

By defining the beam diameter in this way, we impose a clear and reproducible geometric boundary on the beam, independent of its underlying intensity distribution. This provides a consistent and experimentally well-defined measure of beam diameter that is directly comparable across different beam types.

\subsection{Cooler `ring' beam production}
\begin{figure}
    \centering
    \includegraphics[width=\linewidth]{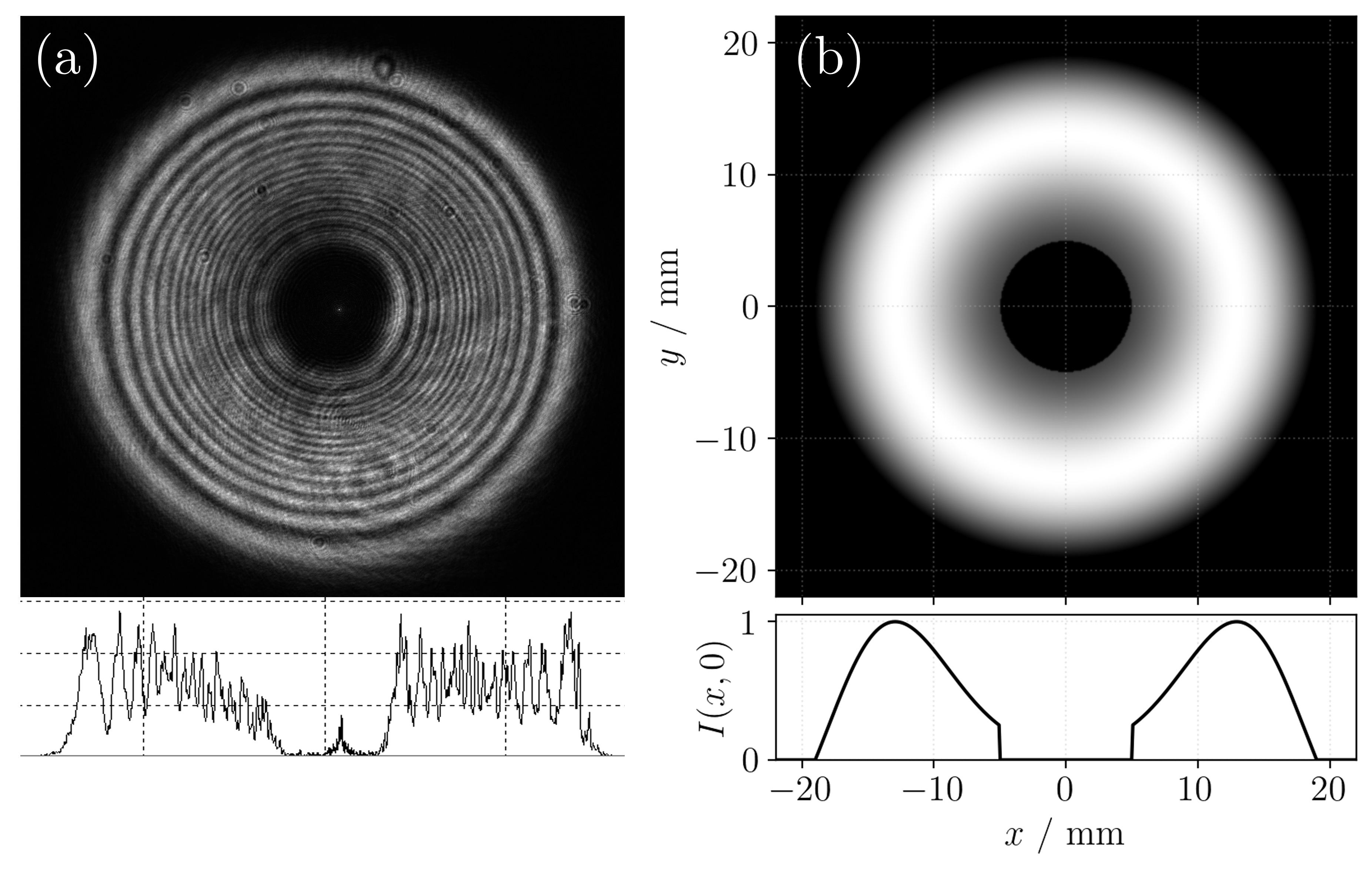}
    \caption{Beam profiles and normalised cross-section intensity distribution for the cooling `ring' beam produced by a pair of axicon lenses, for (a) the experimentally observed beam, and (b) the theoretically calculated profile expected when a clipped Gaussian beam (with waist $10\,\mathrm{mm}$ and clipping radius $14\,\mathrm{mm}$) undergoes a transformation produced by a pair of perfect axicon lenses separated beyond the depth of focus.}
    \label{fig:axicon beam real and sim}
\end{figure}

A pair of axicon lenses is used to transform an initially Gaussian input beam into a hollow, ring-shaped intensity distribution, which we see in our experiment (see fig. \ref{fig:axicon beam real and sim}). The formation of the ring beam can be understood geometrically in terms of ray optics. Each point on the input beam is mapped to a corresponding propagation angle determined by the axicon cone angle. After propagation over a suitable distance, beyond the depth of focus, rays incident at a given initial radius cross to form a surface whose cross-section is an annulus. The second axicon is then placed to `collimate' the profile of the beam. 

The diameter and distribution of this ring is primarily determined by the axicon cone angle and the effective propagation distance between the lenses, allowing straightforward tuning of the trap geometry by adjusting the optical alignment. An iris before the beam manipulation is also included, as this allows for adjustments of the central hole region produced in the intensity profile. The resulting spatial profile seen in the experiment, alongside the theoretical intensity distribution produced with 3D ray optics, is seen in fig \ref{fig:axicon beam real and sim}.

%


\section{Multi-frequency generation}

We identified two convenient methods of generating light with the required spectral properties for this experiment. Full details of these methods are to be made available in a separate publication. However, a very brief description of each is given below. 

In the first method, an acousto-optic frequency shifter is placed inside a very low finesse ($<2$) ring cavity, constructed by coupling light in and out of a circulating pathway with a standard beam splitter. An optical gain medium, in the form of a tapered amplifier, is also added within the ring cavity and used to compensate the majority of the optical losses. On each pass around the ring, the light obtains a frequency shift from the acousto-optic device, experiences substantial power loss from both imperfections and intentional outcoupling, and is then coherently amplified to close to its original power in the optical gain medium. A series of spectral components is thus produced with identical frequency spacing and a constant power scaling between each spectral component and the next, yielding an exponentially decaying optical power as a function of total frequency shift. Since the acousto-optic device always shifts the optical frequency in the same direction, the resulting spectrum is single-sided, with a hard cutoff in one direction. 

\begin{figure}
    \centering
    \includegraphics[width=\linewidth]{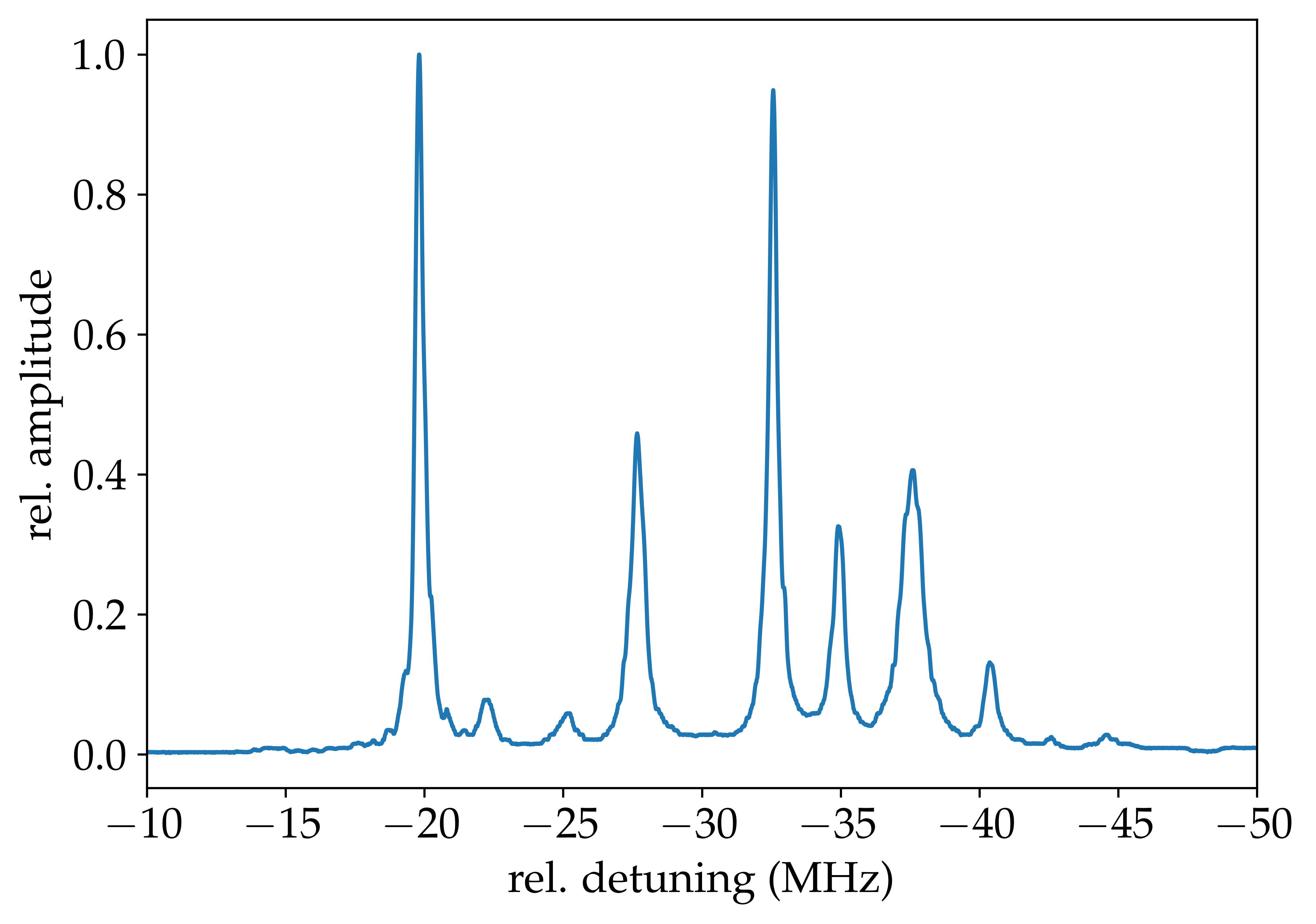}
    \caption{Frequency peaks present in the cooling/trapping MOT beams, measured with a Fabry--P\'erot interferometer, relative to the $F=2 \rightarrow F'=3$ cooler transition of the \textsuperscript{87}Rb D\textsubscript{2} line. The left-most peak at $\sim$--20\,MHz is the `trapping' frequency and exists as a Gaussian distribution across the entirety of the MOT beams; the other peaks are the `cooling' frequencies and exist solely in the axicon ring beam. Whilst the relative amplitude of the cooler frequencies can be a comparison of their relative powers in the beam, the amplitude of the trapping frequency compared to the cooling frequencies is arbitrary due to the nature in which the beams are combined.}
    \label{fig:MF FP peaks}
\end{figure}

The second method, which for practical reasons was used to take the data presented in this manuscript (see fig. \ref{fig:MF FP peaks}), employed a double-passed acousto-optic modulator driven with multiple RF frequencies simultaneously. By placing the retroreflecting mirror for the double pass behind a lens, with the focal plane of the lens matched to the acousto-optic modulator crystal, the different frequency components produced could be recombined into a single, spatially-overlapped beam, despite their different deflection angles in the modulator. This process is very inefficient in optical power, in a best case scenario resulting in a power reduction of at least $1/n$ for the production of $n$ distinct frequency components. For this reason, a tapered amplifier was used after multi-frequency generation, to amplify the resulting output to the power levels needed for the experiment. 

In both methods, a second acousto-optic device was used to partially cancel the central frequency shift of the first. This is not necessary in principle, but was required for practical reasons due to a mismatch between the scale of the frequency spacing desired between components ($\sim$5\,MHz) and the frequency shifts produced by commercial acousto-optic devices (typically $>40$\,MHz).

\section{Simulations}
\label{simulation info}
For simulations varying the magnetic field gradient, the full rubidium-87 atom was modeled, formulating the coupled hyperfine state Hamiltonian for the full $5^2\mathrm{S}_{1/2}$ and $5^2\mathrm{P}_{3/2}$ manifolds of the D\textsubscript{2} line, in the PyLCP software suite \cite{PyLCP}. Laser beams were constructed---a repumper beam tuned on resonance with the $F=1\rightarrow F'=2$ transition, and trapping and cooling beams tuned to $\delta_i$ from the $F=2 \rightarrow F'=3$ transition. The trapping beam detuning was $\delta_\mathrm{trap.}=-3.3\Gamma$, while the detunings for the cooling beams were $\delta_{\mathrm{cool.},i}=(-6\Gamma,\,-7\Gamma,\,-8\Gamma)$ with a weighting distribution (0.23, 0.54, 0.23) of the total power in the cooling beam. This closely followed experimentally observed values at the time.

The repumper and trapping beams had an intensity distribution given by a Gaussian beam profile clipped at $r=1.25w=19\,\mathrm{mm}$, where $w$ is the $1/e^2$ radius. This closely matched the experimentally observed profile of the beams. The cooling beams were modeled with a top hat intensity distribution for simplicity. The intensities within those profiles were calculated to give total powers in one beam $P_\mathrm{repump.}=20\,\mathrm{mW}$, $P_\mathrm{trap.}=36\,\mathrm{mW}$ and $P_\mathrm{cool.}=99\,\mathrm{mW}$ respectively, matching values obtained at the time of the simulation. 

To reduce computational requirements and enable timely evaluation, simulations involving the varying of the beam diameter were modified to a simple two-level system ($F=0 \rightarrow F'=1$). This offered similar trends when compared to the full model and only differed slightly numerically, owing to the difference in Zeeman splittings of the states (e.g., the lower manifold in this model has one single state with no Zeeman splitting present) with different hyperfine Land\'e $g_F$ factors. Again, the trapping beam was set up with a Gaussian beam profile and the cooling beams a top hat beam. No repumper was included, since the manifolds established in the simulation did not require a repumper for cycling transitions (no hyperfine splitting present in the $F=0$ ground state). 

For simulations that match our experiment (solid green and orange lines in fig. \ref{fig:vs beam diam}), the beam diameter was truncated by varying amounts. The trapping beam had a fixed $1/e^2$ beam waist radius of $w=15.2\,\mathrm{mm}$. The central intensity of this beam was calculated to give the full $r=1.25w=19\,\mathrm{mm}$ beam a power of $P_\mathrm{trap.}=19\,\mathrm{mW}$. The top hat cooling beam intensity was calculated to give the total power $P_\mathrm{cool.}=45\,\mathrm{mW}$ for the full $r=19\,\mathrm{mm}$ beam. These values were chosen as they reflected the experimentally observed values at the time. The trapping beam detuning was $\delta_\mathrm{trap.}=-3.3\Gamma$, while the cooling beam detunings were $\delta_{\mathrm{cool.},i}=(-4.5\Gamma,\, -5\Gamma,\, -5.5\Gamma,\, -6\Gamma)$ with a weighting distribution of (0.15, 0.35, 0.35, 0.15) of the total cooling power, closely matching what was experimental observed. The beams were then truncated at varying values to give the truncation diameter, whilst keeping all distributions otherwise the same.

The simulations exploring the continuation of our experiment beyond experimental limitations are displayed as dash-dotted lines in fig. \ref{fig:vs beam diam}. Instead, the $1/e^2$ Gaussian trapping beam waist $w$ was varied whilst keeping the truncation ratio $r=1.25w$ fixed for both the Gaussian trapping beam and top hat cooling beams; the the value $2r$ recorded as beam diameter. To produce a meaningfully comparative simulation, the intensity for each frequency peak present in the cooling top hat beam was matched to be 0.3 times the central intensity of the trapping beam, which was fixed at the value calculated above. This allowed the power to increase for larger beam diameters, offering speculative data from these simulations. Again, the detuning of the trapping beam was $\delta_\mathrm{trap.}=-3.3\Gamma$. Since these simulations were to provide demonstrative insights into the potential gain available from multi-frequency cooling light without experimental limitations, the number of cooling beam detunings was increased, to a total of 10 peaks from $-4\Gamma$ to $-13\Gamma$ spaced every $\Gamma$.

To highlight the further gain available beyond our experimental parameters, we also simulate a regime which is more optimized for larger beam diameters. This is displayed as solid blue and red lines in fig \ref{fig:vs beam diam}. Here, the field gradient is halved (in line with the lower field gradient for larger beam diameters in \cite{Camara2014VLMOT}). Both the cooling and trapping beams are modeled as top hat beams to allow for experimental continuation, and the diameter of the top hat beams recorded. The trapping beam was set to have an intensity $s_0=1$ with a detuning of $\delta_\mathrm{trap.}=-3\Gamma$, whilst the cooling beams were set up to have $s_0=1$ per frequency peak, where like before a total of 10 peaks from $-4\Gamma$ to $-13\Gamma$ spaced every $\Gamma$ are deployed.

These beam profiles were used to form the three counter-propagating orthogonal beam pairs used to construct a typical MOT, with the polarizations tuned in the typical configuration. The beam pairs were established in the $\hat{\mathbf{z}}$-direction, the $(\hat{\mathbf{x}}+\hat{\mathbf{y}})$-direction, and the $(\hat{\mathbf{x}}-\hat{\mathbf{y}})$-direction. A quadrupole magnetic field of the form $\mathbf{B}=b(-x\hat{\mathbf{x}}/2-y\hat{\mathbf{y}}/2 + z\hat{\mathbf{z}})$  was established with a given axial field gradient strength $b$, to complete the full trap parameters.

Combining all of the light--matter interactions available in the system, the governing rate equation was constructed. From this the motion of a single atom was evolved from one end of the trapping region (at a beam radius away from the center of the trap), given characteristics such as it's position in the trap, it's velocity, and consequent forces and resonances with laser beams. The atomic motion was confined to the $\hat{\mathbf{x}}$-direction, whilst maintaining the three-dimensional MOT beam configuration for atomic saturation purposes---this is different from previous simulations \cite{MFMOT_fail} where cross-saturation effects were neglected; it is seen that for the high beam powers used power broadening is in fact significant enough for the orthogonal beams to contribute to the saturation of atoms and affect the loading rate by a factor of up to 1.5. Through iterative numerical integration of the coupled ODE system, the capture velocity $v_c$ is identified as the critical initial velocity at the boundary between the regimes where an atom remains confined within the trapping region and that where it ultimately escapes.

Since atomic flux density scales as the fourth power of velocity for low velocities in a thermal background, we make the approximation $R\propto v_c^4$, where $R$ is the loading rate of the MOT. This is consistent with approximations in the Reif model $v_c\ll v_\mathrm{th}$ \cite{Gibble1992} and loading from a low-pressure uniform background \cite{reif1965fundamentals,steane:1992}. A single free scaling factor, fitted from experimental data and applied identically across all simulations, allows for uncertainties in atomic vapour pressure, how one defines the area of the trapping region etc. \cite{Haw:12,Monroe:1990, Monroe:91}. Additionally, to account for the three-dimensional magnetic field asymmetry produced by the anti-Helmholtz quadrupole coil compared to the one-dimensional motional confinement, a fitted scaling factor (confined to lie between $1/\sqrt{2}$ and $\sqrt{2}$) has been applied uniformly to the magnetic field gradient in all simulations.

To model the steady-state atom number, we obtain an empirical characteristic load time $\tau$ from a fit of eq. (\ref{loadrate}) to experimental data under the simulated conditions or, where they extend beyond the range of our experimental parameters, under the closest available experimental conditions. From the relation $N_\mathrm{ss}=R\tau$ we then obtain values for the steady-state atom number from our capture velocity simulations. It should be noted that this model does not fully capture the physics of density-dependent atomic interactions, and as such should only be seen as a guide to approximate trends in steady state atom number, not an accurate quantitative prediction.


% ...

\newpage

\bibliography{refs}

@article{oppenheim2023gravitationally,
  title={Gravitationally induced decoherence vs space-time diffusion: testing the quantum nature of gravity},
  author={Oppenheim, Jonathan and Sparaciari, Carlo and {\v{S}}oda, Barbara and Weller-Davies, Zachary},
  journal={Nature Communications},
  volume={14},
  number={1},
  pages={7910},
  year={2023},
  publisher={Nature Publishing Group UK London},
doi={https://doi.org/10.1038/s41467-023-43348-2}
}

@article{Howl2021,
  title = {Non-Gaussianity as a Signature of a Quantum Theory of Gravity},
  author = {Howl, Richard and Vedral, Vlatko and Naik, Devang and Christodoulou, Marios and Rovelli, Carlo and Iyer, Aditya},
  journal = {PRX Quantum},
  volume = {2},
  issue = {1},
  pages = {010325},
  numpages = {32},
  year = {2021},
  month = {Feb},
  publisher = {American Physical Society},
  doi = {10.1103/PRXQuantum.2.010325},
  url = {https://link.aps.org/doi/10.1103/PRXQuantum.2.010325}
}

@article{hoffnagle_clock,
author = {Sang Eon Park and Ho Seong Lee and Eun-joo Shin and Taeg Yong Kwon and Sung Hoon Yang and Hyuck Cho},
journal = {J. Opt. Soc. Am. B},
number = {11},
pages = {2595--2602},
publisher = {Optica Publishing Group},
title = {Generation of a slow and continuous cesium atomic beam for an atomic clock},
volume = {19},
month = {Nov},
year = {2002},
url = {https://opg.optica.org/josab/abstract.cfm?URI=josab-19-11-2595},
doi = {10.1364/JOSAB.19.002595},
}

@article{KMF2DMOT,
author = {Jae Hoon Lee and Jongchul Mun},
journal = {J. Opt. Soc. Am. B},
number = {7},
pages = {1415--1420},
publisher = {Optica Publishing Group},
title = {Optimized atomic flux from a frequency-modulated two-dimensional magneto-optical trap for cold fermionic potassium atoms},
volume = {34},
month = {Jul},
year = {2017},
url = {https://opg.optica.org/josab/abstract.cfm?URI=josab-34-7-1415},
doi = {10.1364/JOSAB.34.001415},
}

@article{LiMF2DMOT,
  title = {Enhanced trapping of cold $^{6}\mathrm{Li}$ using multiple-sideband cooling in a two-dimensional magneto-optical trap},
  author = {Li, Kai and Zhang, Dongfang and Gao, Tianyou and Peng, Shi-Guo and Jiang, Kaijun},
  journal = {Phys. Rev. A},
  volume = {92},
  issue = {1},
  pages = {013419},
  numpages = {6},
  year = {2015},
  month = {Jul},
  publisher = {American Physical Society},
  doi = {10.1103/PhysRevA.92.013419},
  url = {https://link.aps.org/doi/10.1103/PhysRevA.92.013419}
}

@Article{portaccelerometer,
author={Tennstedt, B.
and Weddig, N.
and Sch{\"o}n, S.},
title={Improved Inertial Navigation With Cold Atom Interferometry},
journal={Gyroscopy and Navigation},
year={2021},
month={Dec},
day={01},
volume={12},
number={4},
pages={294-307},
issn={2075-1109},
doi={10.1134/S207510872104009X},
url={https://doi.org/10.1134/S207510872104009X}
}

@article{borntest,
author = {Kanthak, Simon and Pahl, Julia and Reiche, Daniel and Krutzik, Markus},
title = {Proposal for a Bose–Einstein Condensate Based Test of Born's Rule Using Light–Pulse Atom Interferometry},
journal = {Advanced Quantum Technologies},
volume = {8},
number = {6},
pages = {2400436},
doi = {https://doi.org/10.1002/qute.202400436},
year = {2025}
}

@Article{fundamentals,
author={O. Buchmueller, J. Ellis and U. Schneider},
title={Large-scale atom interferometry for fundamental physics},
journal={Contemporary Physics},
year={2023},
volume={64},
pages={93-110},
}

@Article{space_fundamentals,
author={Bassi, A.
and Cacciapuoti, L.
and Capozziello, S.
and Dell'Agnello, S.
and Diamanti, E.
and Giulini, D.
and Iess, L.
and Jetzer, P.
and Joshi, S. K.
and Landragin, A.
and Poncin-Lafitte, C. Le
and Rasel, E.
and Roura, A.
and Salomon, C.
and Ulbricht, H.},
title={A way forward for fundamental physics in space},
journal={npj Microgravity},
year={2022},
month={Nov},
day={02},
volume={8},
number={1},
pages={49},
issn={2373-8065},
doi={10.1038/s41526-022-00229-0},
url={https://doi.org/10.1038/s41526-022-00229-0}
}

@article{zeeman_2025,
  title = {Compact dual-beam Zeeman slower for high-flux cold atoms},
  author = {Chen, Chen and Liu, Kejun and Deng, Dezhou and Ma, Shuchang and Zhu, Peng and He, Zhichang and Chen, J.F. and Wu, Xiaoxiao and Chen, Peng},
  journal = {Phys. Rev. Appl.},
  volume = {24},
  issue = {5},
  pages = {054071},
  numpages = {10},
  year = {2025},
  month = {Nov},
  publisher = {American Physical Society},
  doi = {10.1103/17w5-9kqq},
  url = {https://link.aps.org/doi/10.1103/17w5-9kqq}
}

@article{zeeman_asaf,
    author = {Paris-Mandoki, A. and Jones, M. D. and Nute, J. and Wu, J. and Warriar, S. and Hackermüller, L.},
    title = {Versatile cold atom source for multi-species experiments},
    journal = {Review of Scientific Instruments},
    volume = {85},
    number = {11},
    pages = {113103},
    year = {2014},
    month = {11},
    issn = {0034-6748},
    doi = {10.1063/1.4900577},
}

@article{hofnagle_slower,
  title = {Continuous high-flux monovelocity atomic beam based on a broadband laser-cooling technique},
  author = {Zhu, M. and Oates, C. W. and Hall, J. L.},
  journal = {Phys. Rev. Lett.},
  volume = {67},
  issue = {1},
  pages = {46--49},
  numpages = {0},
  year = {1991},
  month = {Jul},
  publisher = {American Physical Society},
  doi = {10.1103/PhysRevLett.67.46},
  url = {https://link.aps.org/doi/10.1103/PhysRevLett.67.46}
}

@article{2dmot_2007,
  title = {Versatile compact atomic source for high-resolution dual atom interferometry},
  author = {M\"uller, T. and Wendrich, T. and Gilowski, M. and Jentsch, C. and Rasel, E. M. and Ertmer, W.},
  journal = {Phys. Rev. A},
  volume = {76},
  issue = {6},
  pages = {063611},
  numpages = {9},
  year = {2007},
  month = {Dec},
  publisher = {American Physical Society},
  doi = {10.1103/PhysRevA.76.063611},
  url = {https://link.aps.org/doi/10.1103/PhysRevA.76.063611}
}

@article{2dmot_2006,
  title = {Realization of an intense cold Rb atomic beam based on a two-dimensional magneto-optical trap: Experiments and comparison with simulations},
  author = {Chaudhuri, Saptarishi and Roy, Sanjukta and Unnikrishnan, C. S.},
  journal = {Phys. Rev. A},
  volume = {74},
  issue = {2},
  pages = {023406},
  numpages = {11},
  year = {2006},
  month = {Aug},
  publisher = {American Physical Society},
  doi = {10.1103/PhysRevA.74.023406},
  url = {https://link.aps.org/doi/10.1103/PhysRevA.74.023406}
}

@article{2dmot_2009,
  title = {High-flux two-dimensional magneto-optical-trap source for cold lithium atoms},
  author = {Tiecke, T. G. and Gensemer, S. D. and Ludewig, A. and Walraven, J. T. M.},
  journal = {Phys. Rev. A},
  volume = {80},
  issue = {1},
  pages = {013409},
  numpages = {12},
  year = {2009},
  month = {Jul},
  publisher = {American Physical Society},
  doi = {10.1103/PhysRevA.80.013409},
  url = {https://link.aps.org/doi/10.1103/PhysRevA.80.013409}
}

@article{twocolour_1994,
author = {Alastair G. Sinclair and Erling Riis and Michael J. Snadden},
journal = {J. Opt. Soc. Am. B},
number = {12},
pages = {2333--2339},
publisher = {Optica Publishing Group},
title = {Improved trapping in a vapor-cell magneto-optical trap with multiple laser frequencies},
volume = {11},
month = {Dec},
year = {1994},
url = {https://opg.optica.org/josab/abstract.cfm?URI=josab-11-12-2333},
doi = {10.1364/JOSAB.11.002333},
}

@phdthesis{timthesis,
    author = {Timothy James},
    title = {Tools and fundamental techniques for
{Bose-Einstein} condensate microscopy},
    school = {University of Sussex},
    year = {2020},
    url = {https://hdl.handle.net/10779/uos.23477252.v1}
}

@article{twocolour_2012,
doi = {10.1088/1674-1056/21/4/043203},
url = {https://doi.org/10.1088/1674-1056/21/4/043203},
year = {2012},
month = {apr},
publisher = {},
volume = {21},
number = {4},
pages = {043203},
author = {C Qiang and L Xin-Yu  and G Kui-Yi  and W Xiao-Rui and C Dong-Min and W Ru-Quan},
title = {Improved atom number with a dual color magneto—optical trap},
journal = {Chinese Physics B}
}

@article{UFG,
doi = {10.1088/0031-8949/91/4/043006},
url = {https://dx.doi.org/10.1088/0031-8949/91/4/043006},
year = {2016},
month = {mar},
publisher = {IOP Publishing},
volume = {91},
number = {4},
pages = {043006},
author = {Törmä, Päivi},
title = {Physics of ultracold Fermi gases revealed by spectroscopies},
journal = {Physica Scripta}
}

@article{
coldmag,
author = {Zhu Ma  and Chengyin Han  and Zhi Tan  and Haihua He  and Shenzhen Shi  and Xin Kang  and Jiatao Wu  and Jiahao Huang  and Bo Lu  and Chaohong Lee },
title = {Adaptive cold-atom magnetometry mitigating the trade-off between sensitivity and dynamic range},
journal = {Science Advances},
volume = {11},
number = {9},
pages = {eadt3938},
year = {2025},
doi = {10.1126/sciadv.adt3938},
URL = {https://www.science.org/doi/abs/10.1126/sciadv.adt3938}}

@article{alphadrift,
doi = {10.1088/1367-2630/aceff6},
url = {https://dx.doi.org/10.1088/1367-2630/aceff6},
year = {2023},
month = {sep},
publisher = {IOP Publishing},
volume = {25},
number = {9},
pages = {093012},
author = {Sherrill, Nathaniel and Parsons, Adam O and Baynham, Charles F A and Bowden, William and Anne Curtis, E and Hendricks, Richard and Hill, Ian R and Hobson, Richard and Margolis, Helen S and Robertson, Billy I and Schioppo, Marco and Szymaniec, Krzysztof and Tofful, Alexandra and Tunesi, Jacob and Godun, Rachel M and Calmet, Xavier},
title = {Analysis of atomic-clock data to constrain variations of fundamental constants},
journal = {New Journal of Physics}
}

@article{f_v_reversal,
  title = {${\mathrm{\ensuremath{\sigma}}}_{+}$-${\mathrm{\ensuremath{\sigma}}}_{\mathrm{\ensuremath{-}}}$ laser-cooling configuration with broadband laser fields: Instability at zero velocity},
  author = {Parkins, A. S. and Zoller, P.},
  journal = {Phys. Rev. A},
  volume = {45},
  issue = {9},
  pages = {R6161--R6164},
  numpages = {0},
  year = {1992},
  month = {May},
  publisher = {American Physical Society},
  doi = {10.1103/PhysRevA.45.R6161},
  url = {https://link.aps.org/doi/10.1103/PhysRevA.45.R6161}
}

@article{gravwav2,
doi = {10.1088/1475-7516/2024/05/027},
url = {https://dx.doi.org/10.1088/1475-7516/2024/05/027},
year = {2024},
month = {may},
publisher = {IOP Publishing},
volume = {2024},
number = {05},
pages = {027},
author = {Baum, Sebastian and Bogorad, Zachary and Graham, Peter W.},
title = {Gravitational wave measurement in the mid-band with atom interferometers},
journal = {Journal of Cosmology and Astroparticle Physics}
}

@article{gravdark,
author = {Badurina, Leonardo  and Buchmueller, Oliver  and Ellis, John  and Lewicki, Marek  and McCabe, Christopher  and Vaskonen, Ville },
title = {Prospective sensitivities of atom interferometers to gravitational waves and ultralight dark matter},
journal = {Philosophical Transactions of the Royal Society A: Mathematical, Physical and Engineering Sciences},
volume = {380},
number = {2216},
pages = {20210060},
year = {2022},
doi = {10.1098/rsta.2021.0060}
}

@Article{gravwav,
author={Beaufils, Quentin
and Sidorenkov, Leonid A.
and Lebegue, Pierre
and Venon, Bertrand
and Holleville, David
and Volodimer, Laurent
and Lours, Michel
and Junca, Joseph
and Zou, Xinhao
and Bertoldi, Andrea
and Prevedelli, Marco
and Sabulsky, Dylan O.
and Bouyer, Philippe
and Landragin, Arnaud
and Canuel, Benjamin
and Geiger, Remi},
title={Cold-atom sources for the Matter-wave laser Interferometric Gravitation Antenna (MIGA)},
journal={Scientific Reports},
year={2022},
month={Nov},
day={08},
volume={12},
number={1},
pages={19000},
issn={2045-2322},
doi={10.1038/s41598-022-23468-3},
url={https://doi.org/10.1038/s41598-022-23468-3}
}

@article{Chu2025,
  title = {Exploring the Dynamical Interplay between Mass-Energy Equivalence, Interactions, and Entanglement in an Optical Lattice Clock},
  author = {Chu, Anjun and Mart\'{\i}nez-Lahuerta, Victor J. and Miklos, Maya and Kim, Kyungtae and Zoller, Peter and Hammerer, Klemens and Ye, Jun and Rey, Ana Maria},
  journal = {Phys. Rev. Lett.},
  volume = {134},
  issue = {9},
  pages = {093201},
  numpages = {9},
  year = {2025},
  month = {Mar},
  publisher = {American Physical Society},
  doi = {10.1103/PhysRevLett.134.093201},
  url = {https://link.aps.org/doi/10.1103/PhysRevLett.134.093201}
}

@article{Kobayashi2024,
  title = {Generation of a precise time scale assisted by a near-continuously operating optical lattice clock},
  author = {Kobayashi, Takumi and Akamatsu, Daisuke and Hosaka, Kazumoto and Hisai, Yusuke and Nishiyama, Akiko and Kawasaki, Akio and Wada, Masato and Inaba, Hajime and Tanabe, Takehiko and Suzuyama, Tomonari and Hong, Feng-Lei and Yasuda, Masami},
  journal = {Phys. Rev. Appl.},
  volume = {21},
  issue = {6},
  pages = {064015},
  numpages = {9},
  year = {2024},
  month = {Jun},
  publisher = {American Physical Society},
  doi = {10.1103/PhysRevApplied.21.064015},
  url = {https://link.aps.org/doi/10.1103/PhysRevApplied.21.064015}
}

@article{troop,
doi = {10.1209/0295-5075/27/8/003},
url = {https://dx.doi.org/10.1209/0295-5075/27/8/003},
year = {1994},
month = {sep},
publisher = {},
volume = {27},
number = {8},
pages = {569},
author = {P. Bouyer and P. Lemonde and M. Ben Dahan and A. Michaud and C. Salomon and J. Dalibard},
title = {An Atom Trap Relying on Optical Pumping},
journal = {Europhysics Letters}
}

@article{collision_loss,
  title = {Collisional losses from a light-force atom trap},
  author = {Sesko, D. and Walker, T. and Monroe, C. and Gallagher, A. and Wieman, C.},
  journal = {Phys. Rev. Lett.},
  volume = {63},
  issue = {9},
  pages = {961--964},
  numpages = {0},
  year = {1989},
  month = {Aug},
  publisher = {American Physical Society},
  doi = {10.1103/PhysRevLett.63.961},
  url = {https://link.aps.org/doi/10.1103/PhysRevLett.63.961}
}

@Article{bichromatic_li,
author={Ilenkov, R. Ya.
and Prudnikov, O. N.
and Kirpichnikova, A. A.
and Taichenachev, A. V.
and Yudin, V. I.},
title={Laser Cooling of Lithium-6 Atoms in a Bichromatic Light Field},
journal={Journal of Experimental and Theoretical Physics},
year={2023},
month={Aug},
day={01},
volume={137},
number={2},
pages={229-238},
issn={1090-6509},
doi={10.1134/S1063776123080058},
url={https://doi.org/10.1134/S1063776123080058}
}

@article{bichromatic_sims,
  title = {Simulation of laser cooling by the bichromatic force},
  author = {Hua, Xiang and Corder, Christopher and Metcalf, Harold},
  journal = {Phys. Rev. A},
  volume = {93},
  issue = {6},
  pages = {063410},
  numpages = {6},
  year = {2016},
  month = {Jun},
  publisher = {American Physical Society},
  doi = {10.1103/PhysRevA.93.063410},
  url = {https://link.aps.org/doi/10.1103/PhysRevA.93.063410}
}

@article{amplifiedDopplercooling,
  title = {Coherent amplification in laser cooling and trapping},
  author = {Freegarde, Tim and Daniell, Geoff and Segal, Danny},
  journal = {Phys. Rev. A},
  volume = {73},
  issue = {3},
  pages = {033409},
  numpages = {9},
  year = {2006},
  month = {Mar},
  publisher = {American Physical Society},
  doi = {10.1103/PhysRevA.73.033409},
  url = {https://link.aps.org/doi/10.1103/PhysRevA.73.033409}
}

@Article{molcool2,
author={Vilas, Nathaniel B.
and Hallas, Christian
and Anderegg, Lo{\"i}c
and Robichaud, Paige
and Winnicki, Andrew
and Mitra, Debayan
and Doyle, John M.},
title={Magneto-optical trapping and sub-Doppler cooling of a polyatomic molecule},
journal={Nature},
year={2022},
month={Jun},
day={01},
volume={606},
number={7912},
pages={70-74},
issn={1476-4687},
doi={10.1038/s41586-022-04620-5},
url={https://doi.org/10.1038/s41586-022-04620-5}
}

@article{molcool1,
doi = {10.1088/1367-2630/aa5ca2},
url = {https://doi.org/10.1088/1367-2630/aa5ca2},
year = {2017},
month = {feb},
publisher = {IOP Publishing},
volume = {19},
number = {2},
pages = {022001},
author = {Truppe, S and Williams, H J and Fitch, N J and Hambach, M and Wall, T E and Hinds, E A and Sauer, B E and Tarbutt, M R},
title = {An intense, cold, velocity-controlled molecular beam by frequency-chirped laser slowing},
journal = {New Journal of Physics}
}

@article{hydrogen,
doi = {10.1088/1367-2630/acf72c},
year = {2023},
month = {sep},
publisher = {IOP Publishing},
volume = {25},
number = {9},
pages = {093038},
author = {Cooper, S F and Rasor, C and Bullis, R G and Brandt, A D and Yost, D C},
title = {Optical deceleration of atomic hydrogen},
journal = {New Journal of Physics}
}

@Article{antihydrogen,
author={Baker, C. J.
and Bertsche, W.
and Capra, A.
and Carruth, C.
and Cesar, C. L.
and Charlton, M.
and Christensen, A.
and Collister, R.
and Mathad, A. Cridland
and Eriksson, S.
and Evans, A.
and Evetts, N.
and Fajans, J.
and Friesen, T.
and Fujiwara, M. C.
and Gill, D. R.
and Grandemange, P.
and Granum, P.
and Hangst, J. S.
and Hardy, W. N.
and Hayden, M. E.
and Hodgkinson, D.
and Hunter, E.
and Isaac, C. A.
and Johnson, M. A.
and Jones, J. M.
and Jones, S. A.
and Jonsell, S.
and Khramov, A.
and Knapp, P.
and Kurchaninov, L.
and Madsen, N.
and Maxwell, D.
and McKenna, J. T. K.
and Menary, S.
and Michan, J. M.
and Momose, T.
and Mullan, P. S.
and Munich, J. J.
and Olchanski, K.
and Olin, A.
and Peszka, J.
and Powell, A.
and Pusa, P.
and Rasmussen, C. {\O}
and Robicheaux, F.
and Sacramento, R. L.
and Sameed, M.
and Sarid, E.
and Silveira, D. M.
and Starko, D. M.
and So, C.
and Stutter, G.
and Tharp, T. D.
and Thibeault, A.
and Thompson, R. I.
and van der Werf, D. P.
and Wurtele, J. S.},
title={Laser cooling of antihydrogen atoms},
journal={Nature},
year={2021},
month={Apr},
day={01},
volume={592},
number={7852},
pages={35-42},
issn={1476-4687},
doi={10.1038/s41586-021-03289-6},
url={https://doi.org/10.1038/s41586-021-03289-6}
}

@article{positronium2,
  title = {Positronium Laser Cooling via the ${1}^{3}S\text{\ensuremath{-}}{2}^{3}P$ Transition with a Broadband Laser Pulse},
  author = {Gl\"oggler, L. T. and Gusakova, N. and Rien\"acker, B. and Camper, A. and Caravita, R. and Huck, S. and Volponi, M. and Wolz, T. and Penasa, L. and Krumins, V. and Gustafsson, F. P. and Comparat, D. and Auzins, M. and Bergmann, B. and Burian, P. and Brusa, R. S. and Castelli, F. and Cerchiari, G. and Ciury\l{}o, R. and Consolati, G. and Doser, M. and Graczykowski, \L{}. and Grosbart, M. and Guatieri, F. and Haider, S. and Janik, M. A. and Kasprowicz, G. and Khatri, G. and K\l{}osowski, \L{}. and Kornakov, G. and Lappo, L. and Linek, A. and Malamant, J. and Mariazzi, S. and Petracek, V. and Piwi\ifmmode \acute{n}\else \'{n}\fi{}ski, M. and Posp\'{\i}\ifmmode \check{s}\else \v{s}\fi{}il, S. and Povolo, L. and Prelz, F. and Rangwala, S. A. and Rauschendorfer, T. and Rawat, B. S. and Rodin, V. and R\o{}hne, O. M. and Sandaker, H. and Smolyanskiy, P. and Sowi\ifmmode \acute{n}\else \'{n}\fi{}ski, T. and Tefelski, D. and Vafeiadis, T. and Welsch, C. P. and Zawada, M. and Zielinski, J. and Zurlo, N.},
  collaboration = {AE\ifmmode \bar{g}\else \={g}\fi{}IS Collaboration},
  journal = {Phys. Rev. Lett.},
  volume = {132},
  issue = {8},
  pages = {083402},
  numpages = {7},
  year = {2024},
  month = {Feb},
  publisher = {American Physical Society},
  doi = {10.1103/PhysRevLett.132.083402},
  url = {https://link.aps.org/doi/10.1103/PhysRevLett.132.083402}
}

@Article{positronium1,
author={Shu, K.
and Tajima, Y.
and Uozumi, R.
and Miyamoto, N.
and Shiraishi, S.
and Kobayashi, T.
and Ishida, A.
and Yamada, K.
and Gladen, R. W.
and Namba, T.
and Asai, S.
and Wada, K.
and Mochizuki, I.
and Hyodo, T.
and Ito, K.
and Michishio, K.
and O'Rourke, B. E.
and Oshima, N.
and Yoshioka, K.},
title={Cooling positronium to ultralow velocities with a chirped laser pulse train},
journal={Nature},
year={2024},
month={Sep},
day={01},
volume={633},
number={8031},
pages={793-797},
issn={1476-4687},
doi={10.1038/s41586-024-07912-0},
url={https://doi.org/10.1038/s41586-024-07912-0}
}

@article{MFMOT_fail,
  title = {Experimental and theoretical study of the vapor-cell Zeeman optical trap},
  author = {Lindquist, K. and Stephens, M. and Wieman, C.},
  journal = {Phys. Rev. A},
  volume = {46},
  issue = {7},
  pages = {4082--4090},
  numpages = {0},
  year = {1992},
  month = {Oct},
  publisher = {American Physical Society},
  doi = {10.1103/PhysRevA.46.4082},
  url = {https://link.aps.org/doi/10.1103/PhysRevA.46.4082}
}

@article{Gibble1992,
author = {Kurt E. Gibble and Steven Kasapi and Steven Chu},
journal = {Opt. Lett.},
number = {7},
pages = {526--528},
publisher = {Optica Publishing Group},
title = {Improved magneto-optic trapping in a vapor cell},
volume = {17},
month = {Apr},
year = {1992},
url = {https://opg.optica.org/ol/abstract.cfm?URI=ol-17-7-526},
doi = {10.1364/OL.17.000526},
}

@article{Li_beam_MOT,
  title = {Enhanced loading of a magneto-optic trap from an atomic beam},
  author = {Anderson, Brian P. and Kasevich, Mark A.},
  journal = {Phys. Rev. A},
  volume = {50},
  issue = {5},
  pages = {R3581--R3584},
  numpages = {0},
  year = {1994},
  month = {Nov},
  publisher = {American Physical Society},
  doi = {10.1103/PhysRevA.50.R3581},
  url = {https://link.aps.org/doi/10.1103/PhysRevA.50.R3581}
}

@article{Meng2024,
  title = {Closed-loop dual-atom-interferometer inertial sensor with continuous cold atomic beams},
  author = {Meng, Zhi-Xin and Yan, Pei-Qiang and Wang, Sheng-Zhe and Li, Xiao-Jie and Xue, Hong-Bo and Feng, Yan-Ying},
  journal = {Phys. Rev. Appl.},
  volume = {21},
  issue = {3},
  pages = {034050},
  numpages = {9},
  year = {2024},
  month = {Mar},
  publisher = {American Physical Society},
  doi = {10.1103/PhysRevApplied.21.034050},
  url = {https://link.aps.org/doi/10.1103/PhysRevApplied.21.034050}
}

@article{solvent_qsim,
  title = {Leveraging analog quantum computing with neutral atoms for solvent configuration prediction in drug discovery},
  author = {D'Arcangelo, Mauro and Henry, Louis-Paul and Henriet, Lo\"{\i}c and Loco, Daniele and Gouraud, Nicola\"{\i} and Angebault, Stanislas and Sueiro, Jules and For\^et, J\'er\^ome and Monmarch\'e, Pierre and Piquemal, Jean-Philip},
  journal = {Phys. Rev. Res.},
  volume = {6},
  issue = {4},
  pages = {043020},
  numpages = {18},
  year = {2024},
  month = {Oct},
  publisher = {American Physical Society},
  doi = {10.1103/PhysRevResearch.6.043020},
  url = {https://link.aps.org/doi/10.1103/PhysRevResearch.6.043020}
}

@article{qsim_spin_interactions,
  title = {Spatially Tunable Spin Interactions in Neutral Atom Arrays},
  author = {Steinert, Lea-Marina and Osterholz, Philip and Eberhard, Robin and Festa, Lorenzo and Lorenz, Nikolaus and Chen, Zaijun and Trautmann, Arno and Gross, Christian},
  journal = {Phys. Rev. Lett.},
  volume = {130},
  issue = {24},
  pages = {243001},
  numpages = {7},
  year = {2023},
  month = {Jun},
  publisher = {American Physical Society},
  doi = {10.1103/PhysRevLett.130.243001},
  url = {https://link.aps.org/doi/10.1103/PhysRevLett.130.243001}
}

@article{Cassens2025,
  title = {Entanglement-Enhanced Atomic Gravimeter},
  author = {Cassens, Christophe and Meyer-Hoppe, Bernd and Rasel, Ernst and Klempt, Carsten},
  journal = {Phys. Rev. X},
  volume = {15},
  issue = {1},
  pages = {011029},
  numpages = {7},
  year = {2025},
  month = {Feb},
  publisher = {American Physical Society},
  doi = {10.1103/PhysRevX.15.011029},
  url = {https://link.aps.org/doi/10.1103/PhysRevX.15.011029}
}

@Article{Georgescu2020,
author={Georgescu, Iulia},
title={{25 years of BEC}},
journal={Nature Reviews Physics},
year={2020},
month={Aug},
day={01},
volume={2},
number={8},
pages={396-396},
issn={2522-5820},
doi={10.1038/s42254-020-0211-7},
url={https://doi.org/10.1038/s42254-020-0211-7}
}

@article{gravimeter2,
  title={Gravity measurements below 10$^{-9}$ g with a transportable absolute quantum gravimeter},
  author={Menoret, V and Vermeulen, P and Le Moigne, N and Bonvalot, S and Bouyer, P and Landragin, A and Desruelle, B},
  journal={Sci. Rep.},
  volume={8},
  pages={12300},
  year={2018},
  doi = {10.1038/s41598-018-30608-1}
}

@article{MIT,
  title={Magnetic induction imaging with a cold-atom radio frequency magnetometer},
  author={Fregosi, A and Gabbanini, C and Gozzini, S and Lenci, L and Marinelli, C and Fioretti, A},
  journal={Appl. Phys. Lett.},
  volume={117},
  pages={144102},
  year={2020},
}

@article{moptrap,
  title={Trapping of $^{85}${R}b atoms by optical pumping between metastable hyperfine states},
  author={Cooper, N and Freegarde, T},
  journal={J. Phys. B},
  volume={46},
  pages={215003},
  year={2013},
  publisher={APS}
}

@misc{Howl2023,
      title={Gravitationally-induced entanglement in cold atoms}, 
      author={Richard Howl and Nathan Cooper and Lucia Hackermüller},
      year={2023},
      eprint={2304.00734},
      archivePrefix={arXiv},
      primaryClass={quant-ph},
      url={https://arxiv.org/abs/2304.00734}, 
}

@article{Robinson2024,
author={Robinson, John M. and Miklos, Maya and Tso, Yee Ming and Kennedy, Colin J. and Bothwell, Tobias and Kedar, Dhruv and Thompson, James K. and Ye, Jun},
title={Direct comparison of two spin-squeezed optical clock ensembles at the 10-17 level},
journal={Nature Physics},
year={2024},
month={Feb},
day={01},
volume={20},
number={2},
pages={208-213},
doi={10.1038/s41567-023-02310-1},
url={https://doi.org/10.1038/s41567-023-02310-1},
}

@article{Sinclair1994,
author = {Alastair G. Sinclair and Erling Riis and Michael J. Snadden},
journal = {J. Opt. Soc. Am. B},
number = {12},
pages = {2333--2339},
publisher = {Optica Publishing Group},
title = {Improved trapping in a vapor-cell magneto-optical trap with multiple laser frequencies},
volume = {11},
month = {Dec},
year = {1994},
url = {https://opg.optica.org/josab/abstract.cfm?URI=josab-11-12-2333},
doi = {10.1364/JOSAB.11.002333},
}

@article{Laudat2018,
doi = {10.1088/1367-2630/aacf1e},
url = {https://doi.org/10.1088/1367-2630/aacf1e},
year = {2018},
month = {jul},
publisher = {IOP Publishing},
volume = {20},
number = {7},
pages = {073018},
author = {Laudat, Théo and Dugrain, Vincent and Mazzoni, Tommaso and Huang, Meng-Zi and Alzar, Carlos L Garrido and Sinatra, Alice and Rosenbusch, Peter and Reichel, Jakob},
title = {Spontaneous spin squeezing in a rubidium BEC},
journal = {New Journal of Physics}
}

@article{Muessel2014,
  title = {Scalable Spin Squeezing for Quantum-Enhanced Magnetometry with Bose-Einstein Condensates},
  author = {Muessel, W. and Strobel, H. and Linnemann, D. and Hume, D. B. and Oberthaler, M. K.},
  journal = {Phys. Rev. Lett.},
  volume = {113},
  issue = {10},
  pages = {103004},
  numpages = {5},
  year = {2014},
  month = {Sep},
  publisher = {American Physical Society},
  doi = {10.1103/PhysRevLett.113.103004},
  url = {https://link.aps.org/doi/10.1103/PhysRevLett.113.103004}
}

@ARTICLE{Chew2022,
  title     = "Ultrafast energy exchange between two single Rydberg atoms on a
               nanosecond timescale",
  author    = "Chew, Y and Tomita, T and Mahesh, T P and Sugawa, S and de
               L{\'e}s{\'e}leuc, S and Ohmori, K",
  journal   = "Nat. Photonics",
  publisher = "Springer Science and Business Media LLC",
  volume    =  16,
  number    =  10,
  pages     = "724--729",
  month     =  oct,
  year      =  2022,
  copyright = "https://creativecommons.org/licenses/by/4.0"
}

@ARTICLE{Hansel2001,
  title     = "{Bose-Einstein} condensation on a microelectronic chip",
  author    = "H{\"a}nsel, W and Hommelhoff, P and H{\"a}nsch, T W and Reichel,
               J",
  journal   = "Nature",
  publisher = "Springer Science and Business Media LLC",
  volume    =  413,
  number    =  6855,
  pages     = "498--501",
  month     =  oct,
  year      =  2001
}

@article{McGuinness2012,
    author = {McGuinness, Hayden J. and Rakholia, Akash V. and Biedermann, Grant W.},
    title = {High data-rate atom interferometer for measuring acceleration},
    journal = {Applied Physics Letters},
    volume = {100},
    number = {1},
    pages = {011106},
    year = {2012},
    month = {01},
    issn = {0003-6951},
    doi = {10.1063/1.3673845},
}

@article{Buchmueller2023,
author = {Oliver Buchmueller and John Ellis and Ulrich Schneider},
title = {Large-scale atom interferometry for fundamental physics},
journal = {Contemporary Physics},
volume = {64},
number = {2},
pages = {93--110},
year = {2023},
publisher = {Taylor \& Francis},
doi = {10.1080/00107514.2023.2239008},


URL = { 
    
        https://doi.org/10.1080/00107514.2023.2239008
    
    

},
eprint = { 
    
        https://doi.org/10.1080/00107514.2023.2239008
    
    

}

}

@article{qgrav1,
doi = {10.1088/2058-9565/abd83e},
url = {https://dx.doi.org/10.1088/2058-9565/abd83e},
year = {2021},
month = {mar},
publisher = {IOP Publishing},
volume = {6},
number = {2},
pages = {024014},
author = {Tino, Guglielmo M},
title = {Testing gravity with cold atom interferometry: results and prospects},
journal = {Quantum Science and Technology}
}

@article{equivtest0,
  title = {Atom-Interferometric Test of the Equivalence Principle at the ${10}^{\ensuremath{-}12}$ Level},
  author = {Asenbaum, Peter and Overstreet, Chris and Kim, Minjeong and Curti, Joseph and Kasevich, Mark A.},
  journal = {Phys. Rev. Lett.},
  volume = {125},
  issue = {19},
  pages = {191101},
  numpages = {5},
  year = {2020},
  month = {Nov},
  publisher = {American Physical Society},
  doi = {10.1103/PhysRevLett.125.191101},
  url = {https://link.aps.org/doi/10.1103/PhysRevLett.125.191101}
}

@Article{equivtest,
author={He, Meng
and Chen, Xi
and Fang, Jie
and Chen, Qunfeng
and Sun, Huanyao
and Wang, Yibo
and Zhong, Jiaqi
and Zhou, Lin
and He, Chuan
and Li, Jinting
and Zhang, Danfang
and Ge, Guiguo
and Wang, Wenzhang
and Zhou, Yang
and Li, Xiao
and Zhang, Xiaowei
and Qin, Lei
and Chen, Zhiyong
and Xu, Rundong
and Wang, Yan
and Xiong, Zongyuan
and Jiang, Junjie
and Cai, Zhendi
and Li, Kuo
and Zheng, Guo
and Peng, Weihua
and Wang, Jin
and Zhan, Mingsheng},
title={The space cold atom interferometer for testing the equivalence principle in the {C}hina Space Station},
journal={npj Microgravity},
year={2023},
month={Jul},
day={28},
volume={9},
number={1},
pages={58},
issn={2373-8065},
doi={10.1038/s41526-023-00306-y},
url={https://doi.org/10.1038/s41526-023-00306-y}
}

@article{AION,
  title={\uppercase{AION}: an atom interferometer observatory and network},
  author={Badurina, L and Bentine, E and Blas, Diego and Bongs, K and Bortoletto, D and Bowcock, T and Bridges, K and Bowden, W and Buchmueller, O and Burrage, C and others},
  journal={J. Cosmol. Astropart. Phys. \textbf{2020} \hspace{-0.22cm}},
  pages={011},
  doi = {10.1088/1475-7516/2020/05/011},
  year={{2020}},
}

@article{Raab1987,
  title={Trapping of neutral sodium atoms with radiation pressure},
  author={Raab, EL and Prentiss, M and Cable, Alex and Chu, Steven and Pritchard, David E},
  journal={Phys. Rev. Lett.},
  volume={59},
  number={23},
  pages={2631},
  year={1987},
  publisher={APS}
}

@article{Schrinski2023,
  title = {Testing collapse models with Bose-Einstein-condensate interferometry},
  author = {Schrinski, Bj\"orn and Haslinger, Philipp and Schmiedmayer, J\"org and Hornberger, Klaus and Nimmrichter, Stefan},
  journal = {Phys. Rev. A},
  volume = {107},
  issue = {4},
  pages = {043320},
  numpages = {12},
  year = {2023},
  month = {Apr},
  publisher = {American Physical Society},
  doi = {10.1103/PhysRevA.107.043320},
  url = {https://link.aps.org/doi/10.1103/PhysRevA.107.043320}
}

@article{ghirardi1990markov,
  title={Markov processes in Hilbert space and continuous spontaneous localization of systems of identical particles},
  author={Ghirardi, Gian Carlo and Pearle, Philip and Rimini, Alberto},
  journal={Physical Review A},
  volume={42},
  number={1},
  pages={78},
  year={1990},
  publisher={APS}
}

@Article{Bongs2019,
author={Bongs, Kai
and Holynski, Michael
and Vovrosh, Jamie
and Bouyer, Philippe
and Condon, Gabriel
and Rasel, Ernst
and Schubert, Christian
and Schleich, Wolfgang P.
and Roura, Albert},
title={Taking atom interferometric quantum sensors from the laboratory to real-world applications},
journal={Nature Reviews Physics},
year={2019},
month={Dec},
day={01},
volume={1},
number={12},
pages={731-739},
issn={2522-5820},
doi={10.1038/s42254-019-0117-4},
url={https://doi.org/10.1038/s42254-019-0117-4}
}

@article{Monroe:91,
author = {C. Monroe and H. Robinson and C. Wieman},
journal = {Opt. Lett.},
number = {1},
pages = {50--52},
publisher = {Optica Publishing Group},
title = {Observation of the cesium clock transition using laser-cooled atoms in a vapor cell},
volume = {16},
month = {Jan},
year = {1991},
url = {https://opg.optica.org/ol/abstract.cfm?URI=ol-16-1-50},
doi = {10.1364/OL.16.000050},
}

@article{Haw:12,
author = {Magnus Haw and Nathan Evetts and Will Gunton and Janelle Van Dongen and James L. Booth and Kirk W. Madison},
journal = {J. Opt. Soc. Am. B},
number = {3},
pages = {475--483},
publisher = {Optica Publishing Group},
title = {Magneto-optical trap loading rate dependence on trap depth and vapor density},
volume = {29},
month = {Mar},
year = {2012},
url = {https://opg.optica.org/josab/abstract.cfm?URI=josab-29-3-475},
doi = {10.1364/JOSAB.29.000475},
}

@article{Monroe:1990,
  title = {Very cold trapped atoms in a vapor cell},
  author = {Monroe, C. and Swann, W. and Robinson, H. and Wieman, C.},
  journal = {Phys. Rev. Lett.},
  volume = {65},
  issue = {13},
  pages = {1571--1574},
  numpages = {0},
  year = {1990},
  month = {Sep},
  publisher = {American Physical Society},
  doi = {10.1103/PhysRevLett.65.1571},
  url = {https://link.aps.org/doi/10.1103/PhysRevLett.65.1571}
}

@article{PyLCP,
title = {{PyLCP: A Python package for computing laser cooling physics}},
journal = {Computer Physics Communications},
volume = {270},
pages = {108166},
year = {2022},
issn = {0010-4655},
doi = {https://doi.org/10.1016/j.cpc.2021.108166},
url = {https://www.sciencedirect.com/science/article/pii/S0010465521002782},
author = {Stephen Eckel and Daniel S. Barker and Eric B. Norrgard and Julia Scherschligt}
}

@ARTICLE{steane:1992,
       author = {{Steane}, A.~M. and {Chowdhury}, M. and {Foot}, C.~J.},
        title = "{Radiation force in the magneto-optical trap}",
      journal = {Journal of the Optical Society of America B Optical Physics},
         year = 1992,
        month = dec,
       volume = {9},
       number = {12},
        pages = {2142-2158},
          doi = {10.1364/JOSAB.9.002142},
       adsurl = {https://ui.adsabs.harvard.edu/abs/1992JOSAB...9.2142S}
}

@book{reif1965fundamentals,
  author    = {Reif, F.},
  title     = {Fundamentals of Statistical and Thermal Physics},
  publisher = {McGraw-Hill},
  address   = {New York},
  year      = {1965}
}

@article{Camara2014VLMOT,
  title = {Scaling behavior of a very large magneto-optical trap},
  author = {Camara, A. and Kaiser, R. and Labeyrie, G.},
  journal = {Phys. Rev. A},
  volume = {90},
  issue = {6},
  pages = {063404},
  numpages = {8},
  year = {2014},
  month = {Dec},
  publisher = {American Physical Society},
  doi = {10.1103/PhysRevA.90.063404},
  url = {https://link.aps.org/doi/10.1103/PhysRevA.90.063404}
}

@ARTICLE{Cheinetinfsens,
  author={Cheinet, Patrick and Canuel, Benjamin and Pereira Dos Santos, Franck and Gauguet, Alexandre and Yver-Leduc, Florence and Landragin, Arnaud},
  journal={IEEE Transactions on Instrumentation and Measurement}, 
  title={Measurement of the Sensitivity Function in a Time-Domain Atomic Interferometer}, 
  year={2008},
  volume={57},
  number={6},
  pages={1141-1148},
  doi={10.1109/TIM.2007.915148}}

@article{Xuecontiousint,
    author = {Xue, Hongbo and Feng, Yanying and Chen, Shu and Wang, Xiaojia and Yan, Xueshu and Jiang, Zhikun and Zhou, Zhaoying},
    title = {A continuous cold atomic beam interferometer},
    journal = {Journal of Applied Physics},
    volume = {117},
    number = {9},
    pages = {094901},
    year = {2015},
    month = {03},
    issn = {0021-8979},
    doi = {10.1063/1.4913711},

}

@article{Duttacontinuous,
   title={Continuous Cold-Atom Inertial Sensor with 1 nrad/sec Rotation Stability},
   volume={116},
   ISSN={1079-7114},
   DOI={10.1103/physrevlett.116.183003},
   number={18},
   journal={Physical Review Letters},
   publisher={American Physical Society (APS)},
   author={Dutta, I. and Savoie, D. and Fang, B. and Venon, B. and Garrido Alzar, C. L. and Geiger, R. and Landragin, A.},
   year={2016},
   month=may }

@article{GallagherPritchard1989,
  title = {Exoergic collisions of cold {Na}$^*$-{Na}},
  author = {Gallagher, Alan and Pritchard, David E.},
  journal = {Phys. Rev. Lett.},
  volume = {63},
  issue = {9},
  pages = {957--960},
  numpages = {0},
  year = {1989},
  month = {Aug},
  publisher = {American Physical Society},
  doi = {10.1103/PhysRevLett.63.957},
  url = {https://link.aps.org/doi/10.1103/PhysRevLett.63.957}
}

@article{Hoffman1994collisional_losses,
author = {Hoffmann, Dominik and Feng, P. and Walker, T.},
year = {1994},
month = {05},
pages = {712-720},
title = {Measurements of Rb trap-loss collision spectra},
volume = {11},
journal = {Journal of the Optical Society of America B},
doi = {10.1364/JOSAB.11.000712}
}

@article{STARKEY2019labscript,
author = "Philip Thomas Starkley",
title = "{A software framework for control and automation of precisely timed experiments}",
year = "2019",
journal = {},
month = "7",
url = "https://bridges.monash.edu/articles/thesis/A_software_framework_for_control_and_automation_of_precisely_timed_experiments/8637200",
doi = "10.26180/8637200.v1"
}

@article{BILLINGTON2019labscript,
author = "Christopher James Billington",
title = "{State-dependent forces in cold quantum gases}",
journal = {},
year = "2019",
month = "4",
url = "https://bridges.monash.edu/articles/thesis/State-dependent_forces_in_cold_quantum_gases/7264406",
doi = "10.26180/5bd68acaf0696"
}

@article{10.1063/1.4817213labscript,
    author = {Starkey, P. T. and Billington, C. J. and Johnstone, S. P. and Jasperse, M. and Helmerson, K. and Turner, L. D. and Anderson, R. P.},
    title = {A scripted control system for autonomous hardware-timed experiments},
    journal = {Review of Scientific Instruments},
    volume = {84},
    number = {8},
    pages = {085111},
    year = {2013},
    month = {08},
    issn = {0034-6748},
    doi = {10.1063/1.4817213},
    url = {https://doi.org/10.1063/1.4817213},

}

\end{document}